\documentclass[12pt,letterpaper]{article}
\usepackage[usenames, dvipsnames]{xcolor}
\usepackage{jcapmod}

\usepackage[margin=2.5cm]{geometry}

\usepackage{tocloft}
\usepackage[]{todonotes}
\usepackage{bbold}
\usepackage{graphicx}
\usepackage{verbatim}
\usepackage{mathrsfs}
\usepackage{slashed}
\usepackage{graphicx}
\usepackage{caption}
\usepackage{subcaption}
\usepackage{bbm}
\usepackage{mathrsfs}

\usepackage{mathrsfs}
\usepackage{relsize}

\usepackage{bbm} 					%%% Blackboard Bold math symbols
\usepackage{slashed} 				%%% Feynman slash
\usepackage{graphicx}				%%% Graphics
\usepackage{subcaption}			%%% Subcaptions for subfigures
\usepackage{psfrag}				%%% Editing EPS files in TeX
\usepackage{tensor}				%%% The \indices command
\usepackage{fouridx}				%%% Arbitrary subscripts/superscripts
\usepackage{bm}					%%% Bold Math symbols
\usepackage{mdframed}				%%% Box for \aside
\usepackage{multirow}				%%% More flexible tables
\usepackage{soul}					%%% Strike-through text
\usepackage{multicol}				%%% Pages with multiple columns

\usepackage{amsmath}
\usepackage{amssymb}
\usepackage{mathtools}

\usepackage{import}

\newcommand*\xoverline[2][0.75]{%
    \sbox{\myboxA}{$\m@th#2$}%
    \setbox\myboxB\null% Phantom box
    \ht\myboxB=\ht\myboxA%
    \dp\myboxB=\dp\myboxA%
    \wd\myboxB=#1\wd\myboxA% Scale phantom
    \sbox\myboxB{$\m@th\overline{\copy\myboxB}$}%  Overlined phantom
    \setlength\mylenA{\the\wd\myboxA}%   calc width diff
    \addtolength\mylenA{-\the\wd\myboxB}%
    \ifdim\wd\myboxB<\wd\myboxA%
       \rlap{\hskip 0.5\mylenA\usebox\myboxB}{\usebox\myboxA}%
    \else
        \hskip -0.5\mylenA\rlap{\usebox\myboxA}{\hskip 0.5\mylenA\usebox\myboxB}%
    \fi}
\makeatother

\newcommand{\ud}{\mathrm{d}}

\newcommand{\gO}[1]{\mathrm{O}\!\left(#1\right)}
\newcommand{\GL}[1]{\mathrm{GL}\!\left(#1\right)}

\DeclareMathOperator{\Span}{span}
\definecolor{cobalt}{RGB}{44, 98, 120}
\definecolor{celadon}{rgb}{0.67, 0.88, 0.69}
\definecolor{dm}{cmyk}{.20, 0, .30, 0}
\definecolor{burgundy}{rgb}{0.5, 0.0, 0.13}
\definecolor{plotBlue}{RGB}{94, 130, 181}

\newcommand{\lab}[1]{{\mathrm{#1}}}

\newcommand{\mb}[1]{{\mathbf{#1}}}
\newcommand{\es}{\hspace{0.5pt}}

\newcommand{\M}{M_\lab{pl}}
\def\be{\begin{equation}}
\def\ee{\end{equation}}
\def\bea{\begin{eqnarray}}
\def\eea{\end{eqnarray}}

\hypersetup{
  colorlinks,
  citecolor=Violet,
  linkcolor=cobalt,
  urlcolor=Blue}

\newif\iffastcompile

\fastcompiletrue

\iffastcompile

\newcommand{\jsi}[1]{}

\else

\newcommand{\jsi}[1]{\todo[color=cobalt!30,size=\scriptsize, bordercolor=cobalt!30, inline]{JS: #1}}

\fi

\newcommand{\email}[1]{\href{mailto:#1}{#1}}
\newcommand{\minus}{{\scalebox {0.75}[1.0]{$-$}}}
\newcommand{\sminus}{{\scalebox {0.5}[0.85]{$-$}}}

\ProvideTextCommandDefault{\Dbar}{%
\leavevmode\lower.5ex\rlap{\hskip-.07em\accent"16}D%
}

\begin{document}
	\newcommand{\main}{.}
\begin{titlepage}

\setcounter{page}{1} \baselineskip=15.5pt \thispagestyle{empty}

\bigskip\

\vspace{2cm}
\begin{center}
{\fontsize{20}{28} \bfseries Instanton Resummation \\[0.4cm] and the Weak Gravity Conjecture  }
 \end{center}
\vspace{1cm}
%\vspace{0.25cm}

\begin{center}
\scalebox{0.95}[0.95]{{\fontsize{14}{30}\selectfont Ben Heidenreich,$^{a}$ Cody Long,$^{b}$ Liam McAllister,$^{c}$}} \vspace{0.35cm}
\scalebox{0.95}[0.95]{{\fontsize{14}{30}\selectfont Tom Rudelius,$^{d}$ and John Stout$^{e}$}}
\end{center}

\begin{center}
\vspace{0.25 cm}
\textsl{$^{a}$Department of Physics, University of Massachusetts, Amherst, MA 01003, USA}\\
\textsl{$^{b}$Department of Physics, Northeastern University, Boston, MA 02115, USA}\\
\textsl{$^{c}$Department of Physics, Cornell University, Ithaca, NY 14853, USA}\\
\textsl{$^{d}$Institute for Advanced Study, Princeton, NJ 08540, USA}\\
\textsl{$^{e}$University of Amsterdam, Amsterdam, the Netherlands}\\

\vspace{0.25cm}
\email{ }
\end{center}

\vspace{1cm}
\noindent

We develop methods for resummation of instanton lattice series.  Using these tools, we investigate the consequences of the Weak Gravity Conjecture for large-field axion inflation.
We find that the Sublattice Weak Gravity Conjecture implies a constraint on the volume of the axion fundamental domain.
However, we also identify conditions under which alignment and clockwork constructions, and a new variant of $N$-flation that we devise, can evade this constraint.
We conclude that some classes of low-energy effective theories of large-field axion inflation are consistent with the strongest proposed form of the Weak Gravity Conjecture, while others are not.

 \vspace{1.1cm}
% \hrule

\bigskip
\noindent\today

\end{titlepage}

\thispagestyle{empty}
\setcounter{page}{2}
\tableofcontents

\clearpage
\pagenumbering{arabic}
\setcounter{page}{2}

\section{Introduction}

Understanding the prospects for large-field inflation in consistent quantum gravity theories is an important question made urgent by the improving measurements of the polarization of the cosmic microwave background \cite{Akrami:2018odb, Abazajian:2016yjj}.

\vskip 3pt
In low-energy effective theories of axion fields, the presence of shift symmetries makes it feasible to compute and control the scalar potential over super-Planckian distances.  At the same time, such theories are plausibly constrained by
quantum-gravitational limits on continuous global symmetries.
For this reason, axion inflation is a promising setting for examining the impact of quantum gravity on large-field inflation.

\vskip 3pt
General quantum gravity arguments suggest that there is a tension between super-Planckian axion periodicities and the computability and convergence of the associated instanton sums \cite{banks:2003sx, Arkanihamed:2006dz}.  In theories with small characteristic periodicities $f \ll M_{\rm{pl}}$, the contributions of higher harmonics to the axion potential are typically suppressed by powers of $e^{-M_{\rm{pl}}/f}$, so the leading---and most readily computable---instantons control the potential, but at the same time the axion fundamental domain is small in Planck units.  Large-field axion inflation is then possible only through axion monodromy \cite{silverstein:2008sg, mcallister:2008hb}, which we will not consider in this work, or through collective excitations of multiple axions \cite{liddle:1998jc}.  On the other hand, for $f \gtrsim M_{\rm{pl}}$ the potential is not typically determined by the instantons carrying minimal axion charges: rather, the sum over the entire charge lattice becomes relevant.  We term this situation a \emph{breakdown of the leading instanton expansion}. It is important to recognize that such a breakdown does not necessarily signal an inconsistency of the theory, and may only present a limitation on our ability to compute.

\vskip 3pt
The Weak Gravity Conjecture (WGC) \cite{Arkanihamed:2006dz} provides a precise incarnation of the apparent tension between large periodicities and the leading instanton expansion.  The WGC asserts that in a consistent effective theory of one or more abelian gauge fields coupled to gravity, there must exist certain extremal or superextremal states, i.e., states whose charge-to-mass ratio equals or exceeds that of an extremal black hole.\footnote{To simplify our language we will write ``superextremal'' instead of ``(super)extremal,'' with the understanding that precisely extremal states are included.}  The corresponding conjecture for axion fields is that there must exist certain instantons whose axion charge-to-action ratio $Q/S$ exceeds $f/M_{\rm{pl}}$, i.e., there must exist instantons with $S \leq Q M_{\rm{pl}}/f$.  We refer to such instantons as \emph{superextremal}.

\vskip 3pt
In its most mild form, the WGC amounts to the \emph{convex hull condition} of \cite{cheung:2014vva}: for any direction $\hat{\mb{Q}}$ in the charge lattice, there exists an instanton of charge $\mb{Q}$ and action $S$ or a collection of instantons of total charge $\mb{Q} = \sum_i \mb{Q}_i$ and total action $S = \sum_i S_i$ satisfying ${S \leq M_{\rm{pl}}|\mb{Q}| /f}$. As a result, it does not place any direct restrictions on low-energy effective field theory: one could imagine that the instantons that satisfy the conjecture have large charges $Q$ and large actions $S$, so that they give negligible, exponentially-suppressed corrections to the instanton potential.

\vskip 3pt
However, stronger forms of the WGC have the potential to place meaningful constraints on models of axion inflation.
In recent years, much work has been devoted to identifying which, if any, is the correct version of the WGC. These different versions are distinguished primarily by which states are required to be superextremal: for example, the lightest charged state, the state with the smallest nonzero charge, etc. Counterexamples to many of the proposed strong forms have been discovered in string theory, but several lines of evidence point to others being true: in particular, all known quantum gravity theories seem to possess not just one, but rather an infinite tower of superextremal particles charged under a given $U(1)$ (see e.g. \cite{Heidenreich:2016aqi, Montero:2016tif, Valenzuela:2016yny, Heidenreich:2017sim, Lee:2018urn, Lee:2018spm, Lee:2019tst}). In the axion context, this translates to an infinite tower of superextremal instantons. These strong forms go under the name Tower WGC (TWGC) \cite{Andriolo:2018lvp}, if one simply requires an infinite tower of superextremal instantons, or Sublattice WGC (sLWGC) \cite{Heidenreich:2015nta, Heidenreich:2016aqi}, if one further requires that these superextremal instantons fill out an entire sublattice of the charge lattice.\footnote{In principle, even the TWGC and sLWGC are not sufficient to constrain low-energy effective theories, as one could suppose that the tower of superextremal instantons begins at a very large charge $Q_{\mathrm{min}} \gg 1$.  However, at present the evidence from string theory weighs against this possibility, and we will assume in this work that $Q_{\mathrm{min}}$ is not large.}

\vskip 3pt
While the evidence for the TWGC and sLWGC has been mounting,
there has not yet been a systematic study of the constraints imposed by these conjectures. In this work, we will explore the constraints imposed by these conjectures on several toy models of axion inflation. We will see that some models, such as isotropic $N$-flation~\cite{dimopoulos:2005ac}, are in tension with the T/sLWGC, provided that the tower of superextremal instantons is sufficiently dense in the charge lattice. But other models, including Kim-Nilles-Peloso (KNP) alignment \cite{kim:2004rp}, clockwork \cite{Choi:2014rja,Choi:2015fiu,Kaplan:2015fuy}, and a modified version of $N$-flation, can be constructed so as to evade all constraints from the T/sLWGC. Furthermore, as has long been appreciated, the bounds imposed by the WGC on axion decay constants do not place meaningful restrictions on models of axion monodromy inflation, as these models do not require super-Planckian decay constants.

\vskip 3pt
Our analysis shows that no proposed strong form of the WGC is strong enough to exclude all low-energy effective theories that support large-field axion inflation.  Even so, it does not offer any insight as to whether models of this sort actually arise in quantum gravity.  Addressing this challenging question may require a direct top-down approach, perhaps in compactifications of string theory.

\vskip 3pt
The organization of this paper is as follows: in \S\ref{sec:AxionsLattices}, we introduce the toy models of axion inflation that we will investigate in subsequent sections, and we review the relevant forms of the WGC. In \S\ref{sec:Constraints}, we derive constraints on axion inflation models, showing in particular that the volume of the fundamental domain of axion field space is bounded by the sLWGC.
It follows that isotropic $N$-flation is incompatible with the sLWGC.  On the other hand, this volume bound does not rule out the possibility of a large effective decay constant in some direction of axion field space, and indeed, we show in \S\ref{sec:Compatible} how to construct models of KNP alignment, clockwork, and modified $N$-flation that are compatible with the sLWGC.
Our conclusions, as well as directions for future research, appear in \S\ref{sec:CONC}.
In Appendix \ref{sec:StringTheory}, we summarize the prospects for ultraviolet completion in string theory of the effective theories discussed herein.
Appendix \ref{sec:saddle} contains more details of the saddle-point computation of \S\ref{sec:Constraints}.

\section{Axions and Lattices}\label{sec:AxionsLattices}
\subsection{Preliminaries}
An axion is a scalar field $\phi$ with an exact discrete shift symmetry $\phi \cong \phi + 2 \pi n f$, $n \in \mathbbm{Z}$. By ``exact,'' we mean not just that the potential is periodic, $V(\phi + 2 \pi f) = V(\phi)$, but moreover that $\phi$ and $\phi + 2 \pi f$ are physically equivalent, i.e., the shift symmetry is gauged. This redundancy has physical consequences, as even Planck-suppressed operators must respect this symmetry.
While this gauged shift symmetry can be spontaneously broken, as in models of axion monodromy, we will only consider the case where it is preserved, as in the original Natural Inflation scenario \cite{freese:1990rb}.

\vskip 3pt
The general two-derivative effective Lagrangian for a single axion takes the form
\begin{equation}
\mathcal{L} = -\frac{1}{2} K(\phi) (\partial \phi)^2 - V(\phi) \,,
\end{equation}
for $K(\phi) > 0$. Since all one-dimensional metrics are flat, we can set $K(\phi) = 1$ by a field redefinition. In this canonically normalized basis, the period $2 \pi f$ takes on physical significance. Note that, although $V(\phi) = V(\phi + 2\pi f)$, $f$ is not fixed by the smallest period of $V(\phi)$. For instance, in theories with extended supersymmetry the potential often vanishes, $V(\phi) = 0$, yet the \emph{axion decay constant} $f$
remains well defined: it is fixed by the period of the discrete gauge symmetry $\phi \cong \phi + 2 \pi f$.\footnote{Similarly, in potentials generated via gaugino condensation, the period of the potential is enhanced by the dual Coxeter number of the gauge group $G$,  so that the inferred decay constant is $f_\lab{eff} = c_2(G) f$. However, this misses the fact that other light states appear as one moves around the axion's fundamental domain $\phi \to \phi + 2 \pi f$, and the \emph{true} ground state energy is invariant under this shift.}

\vskip 3pt
An alternate perspective is provided by the \emph{lattice basis}, $\theta \equiv \phi/f$. In this basis, the two-derivative effective action takes the form
\begin{equation}
\mathcal{L} = -\frac{1}{2} f^2 (\partial \theta)^2 - V(\theta) \,, \mathrlap{\qquad \theta \cong \theta+2 \pi \,.}
\end{equation}
Here the period of the discrete shift symmetry $\theta \cong \theta +2 \pi$ is fixed and the axion decay constant can be read off from the normalization of the kinetic term. The difference between the canonically normalized basis and the lattice basis is analogous to the difference between canonical and holomorphic normalizations for gauge fields. Indeed, there is a direct connection, as we will see below.

\vskip 3pt
In the lattice basis, it is straightforward to generalize to the case of $N$ axions $\theta^i$, each with a discrete shift symmetry $\theta^i \cong \theta^i + 2 \pi$. A general two-derivative effective Lagrangian now takes the form
\begin{equation}
\mathcal{L} = -\frac{1}{2} K_{i j} (\theta) \partial_\mu \theta^i\, \partial^\mu \theta^j - V(\theta^i) \,,\mathrlap{\qquad \theta^i \cong \theta^i+2\pi \,,}
\label{eq:latticebasis}
\end{equation}
where $K_{i j}(\theta)$ is a positive-definite field-space metric.
For simplicity, we will assume that $K_{i j}(\theta)$ is flat, so that we can take it to be constant after a field redefinition. If we then canonically normalize, $\phi^a \equiv f\indices{^a_i} \theta^i$, where $K_{i j} = f\indices{^a_i} {f}_{j a} $, our effective Lagrangian is
\begin{equation}
\mathcal{L} = -\frac{1}{2} \delta_{ab} \,\partial_\mu \phi^b \partial^\mu \phi^b -  V(\phi^a) \,, \mathrlap{\qquad \bm{\phi} \cong \bm{\phi} + 2 \pi\es \mb{f}_{\es\es i}  \,.}
\label{eq:axionLagrangian}
\end{equation}
The $N$ periods $\mb{f}_{\es\es i}$ generate the \emph{period lattice} $\Gamma^\ast = \Span_{\mathbbm{Z}} \{ \mb{f}_{\es\es i} \}$, in terms of which the discrete shift symmetry can be written concisely as
\begin{equation}
\bm{\phi} \cong \bm{\phi} +2 \pi \Gamma^\ast \,.
\end{equation}
 For a single axion, we have $\Gamma^\ast = f \mathbbm{Z}$, and so the period lattice $\Gamma^\ast$ is the appropriate $N$-axion generalization of the axion decay constant.
This lattice can also be thought of as a magnetic charge lattice, since the charges of
codimension-two defects---i.e.,~cosmic strings---satisfy the quantization condition
 \begin{equation}
 \frac{1}{2 \pi} \oint_{S^1}  \ud \bm{\phi} \in \Gamma^\ast \,.
 \end{equation}
We may then define the axion fundamental domain as the field space accessible to the axions modulo periodic identification, which is given by $\mathbbm{R}^N / (2 \pi \Gamma^*)$. Its volume is
\begin{equation}
  \lab{vol}\, \bm{\phi} = |2 \pi \Gamma^* | = (2 \pi)^N | \Gamma^* |\,,
\end{equation}
where $| \Gamma^*|$ denotes the volume of the $N$-torus $\mathbbm{R}^N/\Gamma^*$, also called the unit cell volume of $\Gamma^*$.

\vskip 3pt
It is crucial to note that both the lattice basis $\theta^i$ and the canonically normalized basis $\phi^a$ are ambiguous. In the former case, there is a $\GL{N,\mathbbm{Z}}$ ambiguity $\theta^i \to M\indices{^i_j} \theta^j$ that preserves the periods $\theta^i \cong \theta^i + 2\pi$ but acts non-trivially on $K \to (M^{-1})^\top K M^{-1}$. Thus, it is really a misnomer to refer to ``the'' lattice basis; there are many! Quantities like ``the eigenvalues of $K$ in the lattice basis'' are not well defined for this reason; they are not $\GL{N,\mathbbm{Z}}$ invariant and so the answer depends on which lattice basis we choose.

\vskip 3pt
In a canonically normalized basis this $\GL{N,\mathbbm{Z}}$ ambiguity acts on the periods $\mb{f}_{\es\es i} \to (M^{-1})\indices{^j_i} \mb{f}_{\es\es j}$, but leaves both $\bm{\phi}$ and the period lattice $\Gamma^\ast$ invariant. Instead, there is an~$\gO{N}$ rotational ambiguity $\phi^a \to \Lambda\indices{^a_b} \phi^b$ that acts on both $\bm{\phi}$ and $\Gamma^\ast$. Because it appeals to geometric intuition, and because rotational invariants are much easier to characterize than $\GL{N,\mathbbm{Z}}$ invariants, we will usually find it convenient to work in a canonically normalized basis.

\vskip 3pt
So far, we have only discussed the axion kinetic term in detail. We argued that if the field-space metric is flat, the physical information in the kinetic term can be naturally encoded in the period lattice $\Gamma^\ast$. We now turn to the potential.
The potential must be well-defined on the axion fundamental domain, and so it is necessarily periodic with $V(\bm{\phi} + 2\pi\Gamma^\ast) = V(\bm{\phi})$. In the lattice basis, the most general potential has the form
\begin{equation}
V(\theta) = \sum_{\ell_1, \ldots, \ell_N} Z_{\ell_1, \ldots, \ell_N} e^{i \ell_i \theta^i} \,,
\label{eq:generalpotential}
\end{equation}
for some complex coefficients $Z_{\ell_1, \ldots, \ell_N}$. The potential $V(\theta)$ is real, and so the coefficients must satisfy $Z_{\sminus \ell_1, \ldots, \sminus \ell_N} = Z_{\ell_1, \ldots, \ell_N}^\ast$. Schematically, the coefficients are of the form
\begin{equation}
Z_{\bm \ell} = A_{\bm{\ell}} e^{- S_{\bm{\ell}} + i \delta_{\bm \ell}}\,,
\end{equation}
where we call $S_{\bm{\ell}}$ the action of the instanton of charge $\bm{\ell}$, $A_{\bm{\ell}}$ is some prefactor, and $\delta_{\bm \ell}$ is a phase.
Moving to a canonically normalized basis, $\ell_i \theta^i = \mb{Q} \cdot \bm{\phi}$, where $Q_a = (f^{-1})\indices{^i_a} \ell_i$. Expressed in terms of $\mb{Q}$, the sum over $\ell_1, \ldots, \ell_N$ becomes a sum over the lattice generated by $({\mb{f}^{-1}})^i$. Since $({\mb{f}^{-1}})^i\cdot \mb{f}_{\es j} = \delta^i_j$, this lattice is simply the \emph{charge} lattice $\Gamma$ dual to the period lattice $\Gamma^\ast$, and thus
\begin{equation}
V(\phi) = \sum_{\mb{Q} \in \Gamma} Z_{\mb{Q}} \, e^{i \mb{Q} \cdot \bm{\phi}} \,, \qquad Z_{\sminus \mb{Q}} = Z_{\mb{Q}}^\ast \,.
\end{equation}
Note that the dual of a lattice $\Gamma$ is defined by $\Gamma^\ast \equiv \{\, \mb{y}\,\es |\,\es \forall \, \mb{x} \in \Gamma\es,\,\mb{x}\cdot\mb{y} \in \mathbbm{Z} \,\}$, hence the symmetry ${V(\bm{\phi} + 2\pi\Gamma^\ast) = V(\bm{\phi})}$ is manifest.

\vskip 3pt
Since $\Gamma^\ast$ is the lattice of magnetic charges, it is natural to think of $\Gamma$ as the lattice of electric charges. Indeed, treating the axion $\phi$ as a zero-form gauge potential, we expect zero-dimensional electrically-charged objects with an axion coupling $S = Q \phi(\bm{x}) $. These are nothing but instantons.
To see their effect, let $\lvert \bm{\phi_0}\rangle$ denote the axion eigenstate $\hat{\bm{\phi}}(\mb{x}) \lvert \bm{\phi_0} \rangle = \bm{\phi_0} \lvert \bm{\phi_0} \rangle$.  We form charge eigenstates
\begin{equation}
| \mb{Q} \rangle := \frac{1}{|2 \pi \Gamma^\ast|} \int_{|2 \pi \Gamma^\ast|} \!\ud^N {\phi} \, e^{i \mb{Q} \cdot \bm{\phi}} | \bm{\phi} \rangle\,,
\end{equation}
where the integral is over the $N$-torus $\mathbbm{R}^N /(2 \pi \Gamma^*)$. By charge conservation, we conclude that an instanton of charge $\mb{Q}_1$ connects $\lvert \mb{Q} \rangle$ with $\lvert \mb{Q} + \mb{Q}_1 \rangle$. In doing so, it generically generates terms in the potential of the form $e^{i \mb{Q}\cdot\bm{\phi}}$.\footnote{Strictly speaking, the potential in (\ref{eq:generalpotential}) is not broken up into an ``instanton'' expansion, but as an expansion in topological sectors. By definition, the potential $V(\bm{\phi})$ is the energy per unit spatial volume, as a function of $\bm{\phi}$, of whatever sector in the theory is generating the axion's potential. This can be written as
\begin{equation}
    V(\bm{\phi}) = \lim_{\mathcal{V} \to \infty} \frac{1}{\mathcal{V}} \sum_{ \mb{Q} \in \Gamma} \langle \mb{0} | \mathcal{H}' | \mb{Q} \rangle\, e^{i \mb{Q} \cdot \bm{\phi}}\,,
\end{equation}
where $\mathcal{H}'$ is the Hamiltonian of that sector, $|\mb{Q} \rangle$ is the ``topological vacuum'' or ``pre-vacuum'' of that sector, and $\mathcal{V}$ is the spatial volume, cf. \cite{Pimentel:2019otp}. Clearly, the transition element $\mathcal{V}^{-1} \langle \mb{0} | \mathcal{H'} |\mb{Q}\rangle$ is the Fourier coefficient~$Z_\mb{Q}$.

In contrast, instantons are generally identified as the \emph{saddles} of a Euclidean path integral. Instantons of all charges will thus contribute to this transition element. For instance, a charge $\mb{Q}'$ instanton can pair with one (or many) instantons with net charge $\mb{Q} - \mb{Q'}$ to contribute to $Z_{\mb{Q}}$.  However, when these instantons are well-localized and have large action, the charge $\mb{Q}$ instanton dominates and one may identify $S_\mb{Q}$ with $Z_\mb{Q} \propto e^{-S_\mb{Q}}$ as the ``instanton action.''

In this paper, we will abuse terminology and refer to the sum over topological sectors as the sum over instantons, but
we emphasize that our expansion of the path integral in terms of semi-classical saddles is only a reliable guide to the dynamics of the theory when those saddles have large action.
\label{footinst}
}

\vskip 3pt
Note that the precise notion of an instanton assumes a large action expansion. In performing a path integral, one must sum over all instantons in the theory, and in the case where the instanton actions are small, the distinction between two instantons of charge $\mb{Q}$ and one instanton of charge $2 \mb{Q}$ becomes unclear. However, the sum over distinct topological sectors labeled by the charge $\mb{Q}$ still makes sense, and in some cases one can explicitly resum the instantons to obtain sensible results.  Resummation of this sort plays an important role in the study of dualities (see, for instance \cite{Candelas:1990rm, Seiberg:1994rs,Nekrasov:2002qd}).

\subsection{Sublattice Weak Gravity Conjecture}\label{ssec:sLWGC}

The mildest version of the Weak Gravity Conjecture (WGC) holds that, in any $d$-dimensional abelian gauge theory coupled to gravity with gauge coupling $e$, there must exist a \emph{superextremal state}, i.e., a state whose charge-to-mass ratio is greater than or equal to that of an extremal black hole \cite{Arkanihamed:2006dz},
\begin{equation}
|e q|/m \geq (|Q|/M)_{\rm ext} \sim 1/M_{\rm pl; d}^{(d-2)/2}\,.
\end{equation}
The WGC is supported by numerous examples in string theory, and there are no known counterexamples.  However, this mild form of the WGC suffers from inconsistencies: it is not necessarily preserved under Higgsing \cite{Heidenreich:2017sim} (see also \cite{Saraswat:2016eaz, Furuuchi:2017upe}) or dimensional reduction \cite{Heidenreich:2015nta}.
Thus, an effective theory resulting from either of these procedures will still satisfy the WGC only if some stronger condition is imposed on the original theory.  In the case of dimensional reduction, the low-energy theory obeys the WGC if one demands that the original theory satisfies the \emph{Sublattice Weak Gravity Conjecture} (sLWGC) \cite{Heidenreich:2016aqi}: for any gauge theory coupled to gravity with charge lattice $\Gamma$, there exists a sublattice $\Gamma_{\rm ext} \subset \Gamma$ of finite index such that for each $\mb{q} \in \Gamma_{\rm ext}$, there is a (possibly unstable) superextremal particle of charge $\mb{q}$.

\vskip 3pt
The sLWGC is satisfied in toroidal orbifold compactifications of heterotic/type II string theory, and it follows (at tree level) from modular invariance in perturbative string theory \cite{Heidenreich:2016aqi}.\footnote{See also \cite{Montero:2016tif} for a similar argument, interpreted in the context of AdS$_3$/CFT$_2$.}  It is also related to the phenomenon of ``gauge-gravity unification" studied in \cite{Heidenreich:2017sim}.

\vskip 3pt
In this paper, we apply the WGC and sLWGC to four-dimensional theories of axions.  For this, we must generalize the WGC from one-form gauge fields to zero-form axions, replacing charged particles with charged instantons, the gauge coupling $e$ with the inverse of the axion decay constant $f$, and the particle mass $m$ with the instanton action $S$.  Thus, for a four-dimensional theory of a single axion with decay constant $f$, the generalized WGC implies the existence of an instanton of charge $Q$ whose action $S$ satisfies
\begin{equation}
\frac{|Q|}{fS} \gtrsim \frac{1}{M_{\rm pl}}\,.
\label{eq:WGC0form}
\end{equation}
Here and henceforth we restrict ourselves to $d=4$, so $M_{\rm pl}\equiv M_{\rm pl;4}$.  This generalized zero-form version of the WGC follows essentially from na\"ive dimensional analysis, but it can be given a more rigorous justification by relating to the one-form version via dimensional reduction \cite{Heidenreich:2015nta} or T-duality (within the context of string theory) \cite{Brown:2015iha}.  The inequality (\ref{eq:WGC0form}) is not precise in the zero-form case, as there is no exact analog of an ``extremal black hole'' in this case (see, e.g.,~\cite{Montero:2015ofa,Heidenreich:2015nta,Hebecker:2016dsw,Hebecker:2018ofv,Hertog:2018kbz,Hebecker:2019vyf} for some attempts in this direction).  However, for the purposes of this paper, we will take ``superextremal" instantons to be those satisfying a sharp bound, $|Q|/(f S) \geq 1/M_{\rm pl}$. Any order-one factors thereby omitted will not qualitatively affect our results.

\vskip 3pt
As discussed above, the precise notion of an instanton, and hence the precise statement of the sLWGC, breaks down when the instanton action $S$ becomes of order unity. In some cases, one can still make precise statements even in the small action limit by relating the axion version of the WGC to the WGC for higher-form objects. For instance, in the context of extranatural inflation \cite{Arkanihamed:2003wu}, the WGC for axions in four dimensions is related to the WGC for one-form gauge fields in five dimensions, which makes sense even when the four-dimensional instanton actions are small \cite{Delafuente:2014aca, Heidenreich:2015wga}. More generally, however, it is not entirely clear why the axion version of the sLWGC should be true, or what form the conjecture should take outside of the large action limit.

\vskip 3pt
The implications of the mild WGC for models of axion inflation have been studied extensively (see e.g. \cite{Arkanihamed:2006dz, rudelius:2015xta, Delafuente:2014aca, Montero:2015ofa, Brown:2015iha, Brown:2015lia, junghans:2015hba, Bachlechner:2015qja,Heidenreich:2015wga,Hebecker:2015rya, Hebecker:2015zss,Ibanez:2015fcv} and references therein).  In this work, we turn our attention to the implications of the sLWGC, which is a much stronger restriction on axion theories.  Generalized to the theory of zero-form axions in (\ref{eq:axionLagrangian}) with charge lattice $\Gamma$, the sLWGC holds that there must exist a sublattice $\Gamma_{\rm ext} \subset \Gamma$ of finite index such that for each $\bm{\ell} \in \Gamma_{\rm ext}$, there is an instanton of charge $\bm{\ell}$ satisfying
\begin{equation}\label{eq:defofslwgc}
|Q_a| \equiv |(f^{-1})^i_{\; a} \ell_i | \geq M^{-1}_{\rm pl}\,,
\end{equation}
where $|Q_a| \equiv (Q_1^2+...+Q_N^2)^{1/2}$ is simply the Euclidean norm of $Q_a$.

\vskip 3pt
An important question for our purposes is how sparse the sublattice $\Gamma_{\rm ext}$ is allowed to be.  If it is too sparse, the sLWGC will place no meaningful restriction on low-energy physics, as the particles satisfying the bound will be heavy objects located far out on the charge lattice.  However, in all known string theory examples the sublattice is not very sparse.

\vskip 3pt
In this paper, we will assume that $\Gamma_{\rm ext} = \Gamma$, so that \emph{every} site in the instanton charge lattice is occupied by a superextremal instanton. A theory with $\Gamma_{\rm ext} = \Gamma$ is said to satisfy the \emph{Lattice Weak Gravity Criterion} (LWGC), and not all theories satisfy this criterion \cite{Arkanihamed:2006dz, Heidenreich:2016aqi}. For our purposes, however, the difference between the LWGC and the sLWGC can be justifiably neglected provided that the $\Gamma_{\rm ext}$ is not too sparse, as it simply introduces further order-one factors into our analysis, which we will ignore.

\section{Inflation Constrained by the LWGC}\label{sec:Constraints}

In this section, we derive constraints on  models of axion inflation that satisfy the sLWGC, as formulated in \eqref{eq:defofslwgc}.
\vskip 3pt
By demanding that the potential is not dominated by high-frequency, LWGC-mandated contributions, we argue in \S\ref{sec:volume} that the volume of the axion fundamental domain must be smaller than the volume of a hypersphere with Planckian radius. This \emph{volume bound} is derived in the continuum limit, in which the charge lattice $\Gamma$ is small enough that lattice sums can be approximated by integrals. We argue that this bound prevents isotropic $N$-flation from realizing controlled super-Planckian displacements, and we extend this conclusion beyond the continuum limit in \S\ref{ssec:isotropic}. Finally, we explore how theories with non-trivial lattices are either constrained by, or skirt past, these constraints in \S\ref{sec:rmt}.

\subsection{The Volume Bound} \label{sec:volume}

In this section, we derive a bound on the volume of the axion fundamental domain. As we will see, this bound will apply in a certain continuum limit, in which the sum over instantons can be approximated as an integral.		
  \vskip 3pt
	 We begin with a Lagrangian, in a canonically normalized basis, of the form,
	 \begin{equation}
	 	\mathcal{L} = -\frac{1}{2} \delta_{ab}\, \partial_\mu \phi^a \, \partial^\mu \phi^b - \Lambda^4 \sum_{\mb{Q} \in \Gamma} Z_{\mb{Q}} \exp\left(i \mb{Q} \cdot \bm{\phi}\right).
		\label{eq:Lag}
	 \end{equation}
	We assume that the potential is randomly drawn from an ensemble, where the magnitudes of the Fourier coefficients are independent and normally distributed, with $|\mb{Q}|$-dependent variances
  \begin{equation}
      \langle |Z_{\mb{Q}}|^2 \rangle = e^{-2 \mu |\mb{Q}|}\,, \label{eq:mu}
  \end{equation}
  and that their phases are independently and randomly distributed over $[0, 2 \pi)$. Here, $|\mb{Q}|$ is the standard Euclidean $2$-norm $|\mb{Q}| = \sqrt{\delta_{ab} Q_a Q_b}$. Intuitively, we should think of the potentials as being generated by the nonperturbative effects of an unknown ultraviolet theory, which we randomize over to reflect our ignorance. As discussed in \S\ref{sec:AxionsLattices}, the size of the axion's fundamental domain is encoded in the structure of the charge lattice and, very roughly, we can think of $\mu |\mb{Q}| \sim \alpha (M_\lab{pl}/f) |\mb{n}|$, where $\alpha \sim \mathcal{O}(1)$ and $\mb{n} \in \mathbbm{Z}^{N}$. This interpretation is exact for isotropic $N$-flation---that is, when $\Gamma^*$ is a cubic  lattice---which we analyze in detail in~\S\ref{ssec:isotropic}.

  \vskip 3pt
Our goal is to characterize the structure of such random potentials, and one measure is to ask: how flat is the potential in a particular direction, $\bar{\mb{e}}$? For almost any direction, we can take $\bar{\mb{e}}$ to be a site in the field space lattice. This ensures that the potential is periodic along that direction, and we can study its structure via the Fourier harmonics,
  \begin{equation}
    \frac{1}{2 \pi} \int_{0}^{2 \pi} \!\ud \psi\, V(\psi \,\bar{\mb{e}}) \, e^{-i n \psi} = \sum_{\mb{Q} \in \Gamma}^{\mb{Q}\cdot \bar{\mb{e}} = n} Z_\mb{Q}\,.
  \end{equation}
We will focus on the \emph{harmonic variances}
  \begin{equation}
    \sigma_n^2(\bar{\mb{e}}) = \sum_{\mb{Q} \in \Gamma}^{\mb{Q} \cdot \bar{\mb{e}} = n} e^{- 2 \mu |\mb{Q}|}\,. \label{eq:harmVariance}
  \end{equation}
  To simplify our notation we typically write $\sigma_n^2(\bar{\mb{e}})$ as $\sigma_n^2$, leaving the dependence on the direction $\bar{\mb{e}}$ implicit.

  \vskip 3pt
The harmonic variance $\sigma_n^2$ measures the extent to which the $n$-th harmonic contributes to the potential along $\bar{\mb{e}}$. Since we assumed that $\langle Z_{\mb{Q}} \rangle = 0$, the $n$-th harmonic is suppressed if $\sigma_n^2 \ll 1$, and otherwise not (absent a statistical fluke).
%~\bh{Is this rephrasing ok?}\lm{yes for me} 
It is worth noting that all information about (\ref{eq:Lag}) along $\bar{\mb{e}} \in \Gamma^*$ is now encoded in both the charge lattice $\Gamma$ and $\bar{\mb{e}}$.

	\vskip 3pt

Intuitively, the LWGC can constrain the size of axionic fundamental domain because, once we make this domain too large, a swarm of instantons contributes to the potential and greatly reduces the distance one can smoothly traverse.  In this limit, the sum (\ref{eq:harmVariance}) receives contributions from an enormous number of sites in the charge lattice, and thus it is not feasible to directly evaluate the sum.  Fortunately, the Poisson summation formula provides a dual representation of a sum over the $N$-dimensional lattice $\Gamma$ in the form of a sum over its dual lattice $\Gamma^*$:
	\begin{align}
		\sum_{\mb{Q} \in \Gamma} f(\mb{Q})&= \frac{1}{|\Gamma|} \sum_{\bar{\mb{Q}} \in \Gamma^*} \int_{\mathbbm{R}^N} \!\!\ud^N Q \, f(\mb{Q}) \exp\left(2 \pi i \mb{Q} \cdot \bar{\mb{Q}}\right) =   \sum_{\bar{\mb{Q}} \in \Gamma^*}  \, \hat{f}(\bar{\mb{Q}})   \,.
		\label{eq:poissonSummation}
	\end{align}
This alternative representation can be an extremely useful way to organize the data in the sum, since if the sum over $f(\mb{Q})$ is very slowly convergent, then the sum over
its Fourier transform $\hat{f}(\bar{\mb{Q}})$ will converge very quickly.

\vskip 3pt	
It will be convenient to work with the generating function of harmonic variances along~$\bar{\mb{e}}$,
	\begin{equation}
		W_{\bar{\mb{e}}}(\psi) = \sum_{\mb{Q} \in \Gamma} e^{-2 \mu |\mb{Q}| + 2 \pi i \psi(\mb{Q} \cdot \bar{\mb{e}})}\,, \label{eq:harmVarGenFun}
	\end{equation}
	such that
	\begin{equation}
		\sigma_n^2 = \int_0^{1} \!\ud \psi\, W_{\bar{\mb{e}}}(\psi)\, e^{-2 \pi i n \psi}\,.
	\end{equation}
	Applying (\ref{eq:poissonSummation}) to (\ref{eq:harmVarGenFun}) yields an alternative representation of the generating function,
	\begin{equation}
		W_{\bar{\mb{e}}}(\psi) = \frac{1}{|\Gamma|} \sum_{\bar{\mb{Q}} \in \Gamma^*} \frac{\mu}{\pi^{(N+3)/2}} \frac{\Gamma(\tfrac{N+1}{2})}{\left[(\bar{\mb{Q}} + \psi \bar{\mb{e}})^2 + (\mu/\pi)^2 \right]^{(N+1)/2}}\,, \label{eq:prHarmVarGenFun}
	\end{equation}
	and thus an alternative representation of the harmonic variances,
	\begin{equation}
		\sigma_n^2 = \frac{2 \mu}{ \pi |\Gamma| |\bar{\mb{e}}|}  \sum_{\bar{\mb{Q}} \in \Gamma^*/\bar{\mb{e}}} \left[ \frac{n^2 |\bar{\mb{e}}|^{-2}}{ \bar{\mb{Q}}_\perp^2+ (\mu/\pi)^2}  \right]^{N/4}\!\! \cos\left(\frac{2 \pi n \, \bar{\mb{e}} \cdot \bar{\mb{Q}}}{\bar{\mb{e}}^2}\right) K_{\tfrac{N}{2}} \left(\frac{2 \pi n}{|\bar{\mb{e}}|} \sqrt{\bar{\mb{Q}}_\perp^2+ (\mu/\pi)^2 }\right)\,. \label{eq:prHarmVar}
	\end{equation}
	Here we have introduced the quantity
	\begin{equation}
		\bar{\mb{Q}}_\perp^2 \equiv \bar{\mb{Q}}^2 - \frac{(\bar{\mb{e}} \cdot \bar{\mb{Q}})^2}{\bar{\mb{e}}^2}\,,
	\end{equation}
	which measures the length of $\bar{\mb{Q}}$'s component perpendicular to the direction $\bar{\mb{e}}$, while $\Gamma^*/\bar{\mb{e}}$ is the sublattice formed by identifying any two lattice vectors $\bar{\mb{Q}}$ and $\bar{\mb{Q}}'$ in $\Gamma^*$ that differ by a multiple of $\bar{\mb{e}}$.
	
  \vskip 3pt
	As is clear from (\ref{eq:poissonSummation}), the origin of the dual lattice point $\bar{\mb{Q}}= 0$ is simply the continuum limit of the sum (\ref{eq:harmVariance}). Interestingly, in this limit the harmonic variances depend only on the length of $|\bar{\mb{e}}|$,
	\begin{equation}
	\sigma_n^2 = \frac{2 \mu}{\pi |\Gamma| |\bar{\mb{e}}|}  \left( \frac{n \pi }{\mu |\bar{\mb{e}}|}\right)^{N/2}\!\!\!\! K_{\tfrac{N}{2}} \left(\frac{2 \mu n}{|\bar{\mb{e}}|}\right) + \dots\,
	\label{eq:continuumVariance}
	\end{equation}
  and not on the actual structure of the lattice, and so (\ref{eq:continuumVariance}) is a \emph{universal result} that applies to all members of this ensemble, as long as the continuum approximation applies. The continuum approximation is valid as long as the shortest vectors in $\Gamma$ are sufficiently small so that many instantons contribute to the original sum (\ref{eq:harmVariance}). As we make these shortest vectors larger, other terms in (\ref{eq:prHarmVar}) will become important and correct the continuum approximation.

  \vskip 3pt
  Since we are primarily interested in the large-$N$ limit, it will be convenient to approximate the modified Bessel function using
  \begin{equation}
    K_\nu (x) \sim \frac{\Gamma(\nu)}{2} \left(\frac{2}{x}\right)^{\nu} e^{-x^2/4 \nu}\Big(1 + \mathcal{O}\big(x/\nu^{3/4}\big)\Big)\, \mathrlap{\qquad \nu \to \infty\,,}
  \end{equation}
  the first term of which, in practice, is a very good approximation to the summand as long as $N \gtrsim 10$.
  %~\bh{For what values of $x$ is this true? $x\ll \nu^{3/4}$?}
  %~\js{just plotting, all values of $x$. Technically, when $x \gg \nu$, I think. }. 
  This allows us to rewrite the harmonic variances (\ref{eq:prHarmVar}) in terms of the volume of the axionic fundamental domain as
  \begin{equation}
    \sigma_n^2 \simeq \frac{2 \mu}{\pi |\bar{\mb{e}}| N}\left( \frac{\lab{vol} \, \bm{\phi}}{\lab{vol}\, D_{N}(2 \mu)} \right) e^{-2 n^2 \mu^2/(N |\bar{\mb{e}}|^2)}  \sum_{\bar{\mb{Q}} \in \Gamma^*/\bar{\mb{e}}} \frac{e^{-2\pi^2 n^2 \bar{\mb{Q}}_\perp^2/(N |\bar{\mb{e}}|^2)}}{[1+\pi^2 \bar{\mb{Q}}_\perp^2/\mu^2]^{ N/2}} \cos\left(\frac{2 \pi n \, \bar{\mb{e}} \cdot \bar{\mb{Q}}}{\bar{\mb{e}}^2}\right)\,, \label{eq:largeNHarmVar}
      \end{equation}
  where we denote the volume of an $N$-ball with radius $2 \mu$ by $\lab{vol}\,D_{N}(2 \mu)$.\footnote{Specifically, $\lab{vol}\, D_{N}(r) = \pi^{N/2} r^N/\Gamma(N/2+1)$.}  From this form, we see that the smallest non-zero values of $(\pi \bar{\mb{Q}}_\perp/\mu)^2$ control the validity of the continuum approximation. Corrections are thus non-universal and depend on the detailed structure of the sublattice $\Gamma^*/{\bar{\mb{e}}}$. However, the continuum approximation becomes more accurate for the higher harmonics since these receive contributions from many more sites in the charge lattice.

  \vskip 3pt
  If we make the axionic fundamental domain smaller and smaller,
  the Poisson-resummed series expansion will eventually break down, and one is better off using the  original expansion in (\ref{eq:harmVariance}). The situation is especially complicated in the intermediate regime in which some directions in the lattice have small $|\mb{Q}|$ while others have large $|\mb{Q}|$: in this case, one must Poisson-resum certain directions in the lattice, but not others. In this work, we will ignore the complications of this intermediate regime and deal exclusively with the two limiting cases: either all $|\mb{Q}|$ are large, so that the sum in (\ref{eq:harmVariance}) can be well-approximated by a small number of terms, or else the continuum approximation is valid. Understanding the intermediate regime requires a significant amount of mathematical machinery, which will be unpacked in a subsequent work~\cite{Resummation}.

  \vskip 3pt
  Working within the continuum limit, the ratios of the harmonic variances are roughly
  \begin{equation}
    \frac{\sigma_{n}^2}{\sigma_1^2} \simeq e^{-2 n^2 \mu^2/(N |\bar{\mb{e}}|^2)}     \simeq 1 \, \quad \text{for}\quad n \ll \frac{\sqrt{N} |\bar{\mb{e}}|}{\mu}\,.
  \end{equation}
 That is, the variances are independent of $n$ and are all roughly equal to one another up to a certain point, after which they decay rapidly. It follows that for an effective decay constant $f_{\text{eff}} = |\bar{\mb{e}}|$, harmonics of wavelength $2 \pi f_{\text{eff}} / n \sim 2 \pi M_{\text{pl}} / \sqrt{N}$ will introduce significant corrections to the inflaton potential. This outcome is the exact opposite of what na\"{i}vely occurs in $N$-flation, in that here increasing $N$ actually \emph{decreases}  the effective field range along the $\bar{\mb{e}}$ direction by a factor of $1/\sqrt{N}$.

 \vskip 3pt
By requiring that the variance in (\ref{eq:largeNHarmVar}) satisfies $\sigma_n^2 \ll 1$ and using the continuum approximation\footnote{The continuum approximation improves for larger $n$ because more instantons contribute to the sum. Thus, even when the approximation fails for small $n$, the volume bound remains valid.} (truncating to the $\bar{\mb{Q}} = 0$ term in the sum) we derive the \emph{volume bound}\footnote{We are assuming that $|\bar{\mb{e}}|$ is not exponentially large in $N$.  If $| \bar{\mb{e}}|$ were exponentially large, then
it would almost certainly be nearly parallel to a lattice vector
$\bar{\mb{e}}'$ whose length is \emph{not} exponentially
large. It is thus unlikely that we can avoid the LWGC-mandated higher harmonics
in this way.}
 \begin{equation}
      \lab{vol}\,\bm{\phi} \lesssim \lab{vol}\, D_{N}(2 \mu)\,, \label{eq:volConstraint}
    \end{equation}
    in the large $N$ limit.
    That is, the requirement that higher harmonics of wavelength $\lesssim \M$ do not contribute significantly to the inflationary potential implies---in theories obeying the LWGC---that the volume of the axionic fundamental domain is bounded by the volume of a \emph{ball} of radius $M_\lab{pl}$ (up to order one factors), and not by a \emph{cube} with Planckian side-length.  Specifically, we find the bound
    \begin{equation}
       \boxed{\frac{\lab{vol}\, \bm{\phi}}{(2 \pi M_\lab{pl})^N} \lesssim \frac{1}{\sqrt{\pi N}} \left[\sqrt{\frac{2 e}{\pi N}} \frac{\mu}{M_\lab{pl}}\right]^{N}} \label{eq:volBound}
    \end{equation}
    in the $N \to \infty$ limit.
%~\bh{Prefactor of $1/\sqrt{\pi N}$ is misleadingly accurate?}~\js{It's fine for me. The $\lesssim$ kinda cancels out any misleading accuracy.}

\begin{figure}
  \begin{center}
    \includegraphics{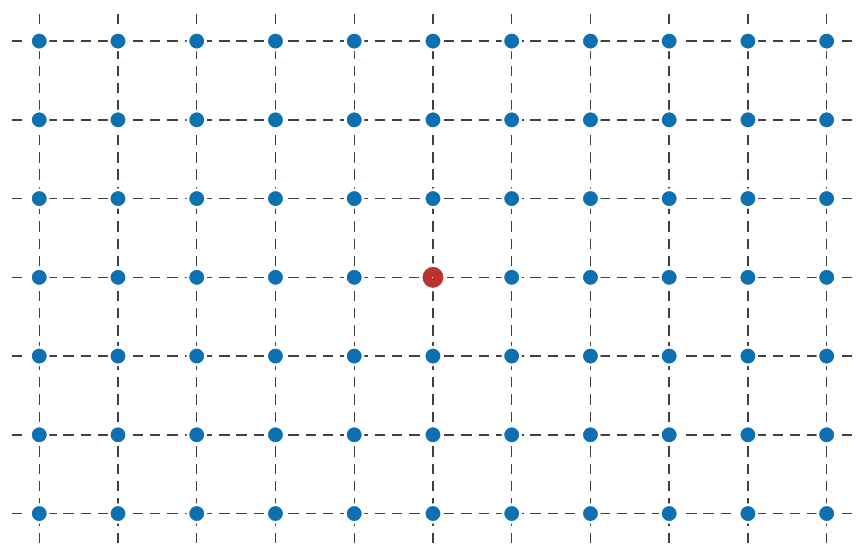}
    \caption{The charge lattice for isotropic $N$-flation. Here, the blue dots indicate charge sites with marginally superextremal instantons, i.e., those that barely satisfy the WGC bound, and the red dot indicates the origin $\mb{Q} = 0$. \label{fig:Nflation}}
  \end{center}
\end{figure}

 \vskip 3pt
    The bound \eqref{eq:volBound} immediately implies that isotropic
$N$-flation (i.e., a model with a cubic charge lattice $\Gamma$, see Figure~\ref{fig:Nflation}) fails in theories obeying the LWGC.  If we violate \eqref{eq:volBound}, the effective field range
will be cut down to $\sim M_{\text{pl}} / \sqrt{N}$ in all directions.
Conversely, if we satisfy \eqref{eq:volBound} by an order-one amount in radius, so that
\begin{equation}
  \lab{vol} \,\bm{\phi} = \lab{vol}\, D_N (r )\,,
\end{equation}
for $r / 2 \mu < 1$ an order-one fraction, then the Poisson-resummed expansion (\ref{eq:prHarmVar}) breaks down and the leading-order instantons will dominate the potential, provided these are not actually dominated by small sublattices of $\Gamma$. This last point in particular suggests that we could get an LWGC-compatible
trans-Planckian field range by making the axion fundamental domain
``pointy,'' with a large field range in one direction balanced by a slightly
smaller field range (the difference is subleading at large $N$) in the other
directions. We show this is true in \S\ref{sec:stretched}.

\vskip 3pt

We caution, however, that the volume constraint \eqref{eq:volBound} should be interpreted as a
\emph{necessary} rather than \emph{sufficient} condition for
inflation compatible with the LWGC: a model may satisfy the volume bound, yet still suffer from higher harmonics that restrict field traversals to be sub-Planckian. For instance, consider a model with $K_{ij} = f_i^2 \delta_{ij} $, and set $f_1 \neq f_2 = f_3 = \ldots = f_N$, with
\begin{equation}
 \prod_{i = 1}^{N} \frac{f_i}{M_\lab{pl}} \lesssim \frac{1}{\sqrt{\pi N}} \left[ \sqrt{\frac{2e}{\pi N}} \frac{\mu}{M_{\lab{pl}}} \right]^N,
\end{equation}
so that the volume bound (\ref{eq:volBound}) is satisfied. One might attempt to generate a super-Planckian field range in the $\bar{\mb{e}} = (1,0,0,...,0)$ direction by taking $f_1 \gg  M_{\lab{pl}}$. However, the LWGC implies that the instantons of charge $(n, 0, \ldots .,0)$, whose action is linearly reduced by the stretching $f_1 \gg M_{\lab{pl}}$, contribute
harmonics of the right size to keep the effective field range sub-Planckian. More generally, even if the potential is flat with
a super-Planckian field range in a certain direction, it remains to be checked
that this is (or is close to) a gradient flow, as required for slow-roll
inflation.\footnote{For instance, any path in an exact moduli space is flat
and can have trans-Planckian length, but exhibiting gradient flows over trans-Planckian distances is more difficult.}
Thus, the prospects for kinetic alignment are not clear from this stage of our analysis, though the volume constraint (\ref{eq:volConstraint}) already gives one non-trivial constraint.

%\newpage

\subsection{Isotropic N-flation}\label{ssec:isotropic}

We have argued above that the LWGC implies the bound \eqref{eq:volBound} on the volume of the axion fundamental domain, which is clearly violated by isotropic $N$-flation models.
The derivation of \eqref{eq:volBound} crucially relied on the continuum approximation, which approximates the sum over the charge lattice as an integral.
In this subsection we will leverage the simplicity of $N$-flation's charge lattice (cf.~Figure~\ref{fig:Nflation}) to compute the harmonic variances to leading order in $1/N$
for all $f$, and thus extend beyond the continuum approximation our finding that isotropic $N$-flation is incompatible with the LWGC.

\vskip 3pt
For simplicity of notation, we will work in the lattice basis defined in (\ref{eq:latticebasis}), such that
 \begin{equation}
    \mathcal{L} = -\frac{1}{2}f^2 \delta_{ij}\, \partial_\mu \theta^i \, \partial^\mu \theta^j - \Lambda^4 \sum_{\bm{\ell} \in \mathbbm{Z}^N} Z_{\bm{\ell}} \exp\left(2 \pi i \bm{\ell} \cdot \bm{\theta}\right).
   \end{equation}
    As in \S\ref{sec:volume} and (\ref{eq:mu}), the Fourier coefficients $Z_{\bm{\ell}}$ are randomly distributed in phase such that $\langle Z_{\bm{\ell}} \rangle = 0$ and
   \begin{equation}
    \langle |Z_{\bm{\ell}}|^2 \rangle = e^{-2 \alpha |\bm{\ell}|}\,,
   \end{equation}
  where $\alpha = \mu / f$ is an $\mathcal{O}(1)$ constant. We are interested in inflating along $\bar{\mb{e}} = (1, 1, \dots, 1)$, as pictured in Figure~\ref{fig:Nflation}. As in the previous section, the variances are defined by
\begin{equation}
    \sigma_n^2 =  \sum_{\bm{\ell} \in \Gamma}^{\bm{\ell} \cdot \bar{\mb{e}}= n} e^{-2 \alpha |\bm{\ell}|} = \sum_s k^{(N, n)}_s e^{- 2 \alpha \sqrt{s}}\,, \label{eq:harmVarS}
\end{equation}
where we have now organized the sum in terms of a multiplicity factor $k_{s}^{(N, n)}$ that counts the number of lattice sites satisfying $\bm{\ell} \cdot \bar{\mb{e}} = \sum_i \ell_i =  n$ and $\bm{\ell}^2 = s$. Specifically, this combinatorial factor counts the number of sites that contribute to the $n$-th harmonic and sit at a distance $\sqrt{s}$ from the origin, and can be written as the sum of multinomial coefficients,
\begin{equation}
  k^{(N, n)}_s = \sum_{\{ r_a \}} \frac{N!}{\prod_a r_a !}\,. \label{eq:multinomialSum}
\end{equation}
Here, $r_a \equiv | \{ \,i\, |\, \ell_i = a \,\} |$ denotes the number of components of $\bm{\ell}$ that are equal to $a$, and the sum is restricted by the conditions
\begin{equation}
  r_a \geq 0 \;, \qquad \sum_a r_a = N \;, \qquad \sum_a a r_a = n \;,
  \qquad \sum_a a^2 r_a = s.
\end{equation}
Computing the harmonic variances in $N$-flation thus reduces to a (relatively) simple combinatorial problem. It is crucial here that the charge lattice is (hyper)-cubic, and general lattices will require more technology \cite{Resummation}.

\vskip 3pt

We will estimate $k^{(N, n)}_s$ in three ways. First, we can approximate it in the continuum limit by replacing the sum over $\mathbbm{Z}^N$ with an integral over $\mathbbm{R}^N$,
\begin{equation}
  \sigma_n^2 \simeq \int_{\mathbbm{R}^N} \!\ud^N \ell\,  e^{- 2 \alpha | \bm{\ell} |}\, \delta \left(\bar{\mb{e}}\cdot \bm{\ell} - n \right) = \frac{\lab{vol} \, \lab{S}^{N-2}}{\sqrt{N}} \int_{n / \sqrt{N}}^{\infty}\!\ud v\,  v
  \left( v^2 - \frac{n^2}{N} \right)^{\frac{N - 3}{2}} \!\!\!e^{- 2 \alpha v}\,. \label{eq:continuumintegral}
\end{equation}
Setting $v= \sqrt{s}$, we see that $k^{(N, n)}_s$ is simply the measure of the integral
 \begin{align}
  k^{(N, n)}_s &\simeq \frac{\lab{vol}\, S^{N - 2}}{\sqrt{N}}  \left[ s - \frac{n^2}{N}
  \right]^{\frac{N - 3}{2}}. \label{eqn:continuumLimit}
  \end{align}
Second, we approximate $k^{(N, n)}_s$ by restricting the sum (\ref{eq:multinomialSum}) to  ``small charge'' instantons, i.e., those with $\ell_i = 0$ and $\pm 1$ for all $i$. The multiplicity is then the multinomial coefficient 
%\bh{Fixed now, I think.}~\js{reintroduced $k_1$ and $k_2$ to avoid small fractions} \bh{Not important, but I don't see any problem with small fractions}
\begin{equation}
  k^{(N, n)}_s \simeq \begin{pmatrix}
    N\\
    k_1 ,\, k_2 ,\, N-s
  \end{pmatrix} =\frac{\Gamma(N+1)}{\Gamma\bigl(k_1 + 1\bigr) \Gamma\bigl(k_2 + 1\bigr) \Gamma(N-s+ 1)}\,,
  \label{eqn:ksLeadingSmallS}
\end{equation}
since we may choose, for each $s$, $k_1 = (s+n)/2$ of the charge vector $\bm{\ell}$'s $N$ entries to be $+1$ and $k_2 = (s-n)/2$ of the entries to be $-1$, and $k_s^{(N,n)}$ counts the number of ways to do this.

\vskip 3pt
Finally, we estimate $k^{(N, n)}_s$ by a saddle point approximation. The details of this computation are too technical to be presented here and can instead be found in Appendix~\ref{sec:saddle}. In the large $N$ limit, the multiplicities are well approximated by
\begin{align}
  \frac{1}{N} \log k_s^{(N, n)} &\simeq \log \theta (w ; q) - \sigma \log q -
  \nu \log w + \mathcal{O}\left(N^{-1} \log N \right)\,. \label{eqn:ksLeadingLargeN}
\end{align}
Here, the Jacobi theta function
\begin{equation}
    \theta(w; q) = \sum_{a \in \mathbbm{Z}} q^{a^2} w^a\,,
\end{equation}
can be interpreted as a thermodynamic free energy, which is a function of the ``chemical potentials'' $w$ and $q$. The values1 of these chemical potentials are determined implicitly in terms of $s$ and $n$ by the equations
\begin{equation}
  \sigma \equiv \frac{s}{N} = \frac{\partial \log \theta(w; q)}{\partial \log  q}\,,
  \end{equation}
and
\begin{equation}
  \nu \equiv \frac{n}{N} = \frac{\partial \log \theta(w; q)}{\partial \log w}\,.
  \end{equation}
These equations can be solved numerically to explicitly evaluate the multiplicity (\ref{eqn:ksLeadingLargeN}) for a given $s$ and $n$.

\begin{figure}
\begin{center}
  \includegraphics{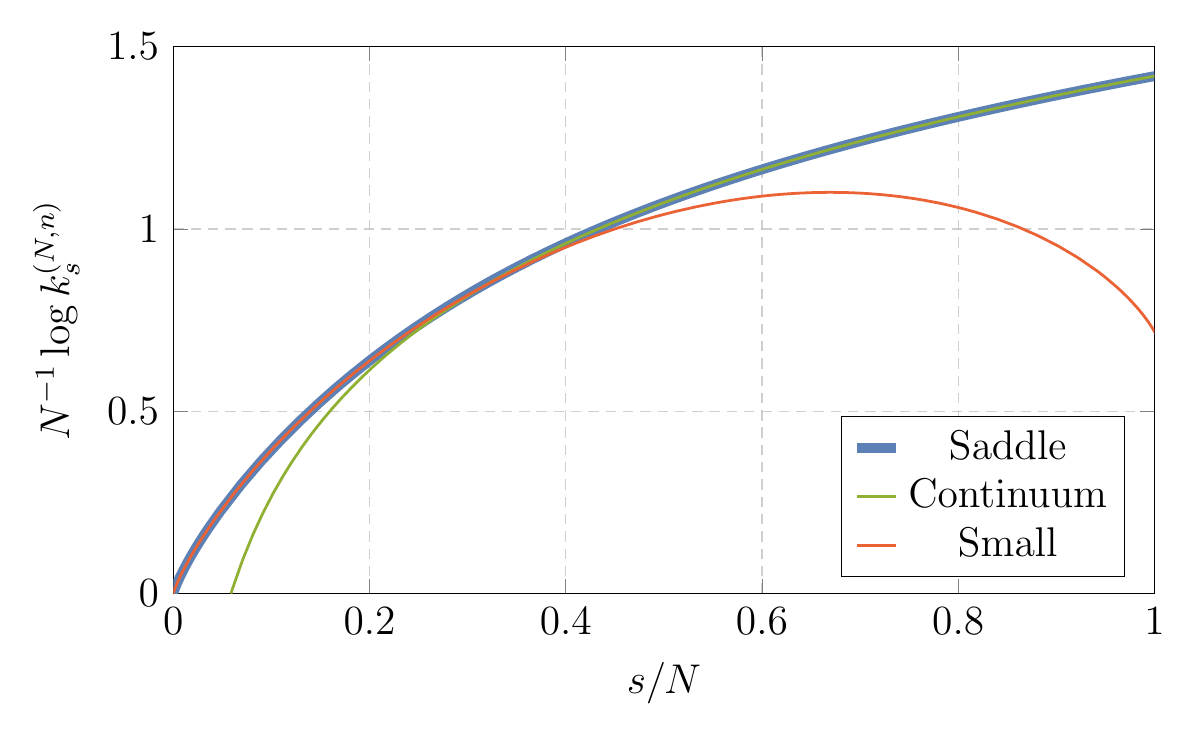}
  \caption{Instanton multiplicities $k_s^{(N, n)}$ computed in the continuum limit (green),  the $\ell_a \in \{ 0, \pm 1 \}$
  approximation (red), and the saddle point approximation (thick, blue) for $\nu = 0$.}
  \label{fig:sigmavsk}
  \end{center}
\end{figure}

\vskip 3pt
Since $f_{\text{eff}} = \sqrt{N} \M$ in an isotropic $N$-flation model, we are interested in the variances $\sigma_n^2$ for $n \lesssim \sqrt{N}$. In the $N \to \infty$ limit, this corresponds to taking $\nu \to 0$. In this limit, we can numerically compute the $k_s^{(N, n)}$ as a function of $\sigma$ under each of the three approximations in (\ref{eqn:continuumLimit}), (\ref{eqn:ksLeadingSmallS}), and (\ref{eqn:ksLeadingLargeN}). The results are plotted in Figure~\ref{fig:sigmavsk}. We see that the continuum limit agrees with the ``exact'' result computed by saddle point approximation for large $s/N$ (far out on the lattice), whereas the sum over instantons of small charge agrees with the saddle point result for small $s/N$. There is an intermediate range in which all three approximations agree, providing strong evidence that our approximations correctly capture the behavior of the isotropic $N$-flation model.

\vskip 3pt

Successful $N$-flation requires that $f \gg \M/\sqrt{N} $, which in turn implies $\alpha \ll \sqrt{N}$. On the other hand, the large-$N$ limit of the volume bound implies that the harmonic variances are exponentially large whenever
\begin{equation}
  \alpha \lesssim \sqrt{\frac{\pi N}{2 e}}\,.
\end{equation}
This means that in the regime of interest, the integral in (\ref{eq:continuumintegral}) is dominated by terms with $|\bm{\ell}| \gg n$, so the continuum limit is valid. This, in turn, means that the volume bound we derived in (\ref{eq:volConstraint}) applies, so we conclude that
\begin{equation}
f_{\text{eff}} = \sqrt{N}f  \lesssim \M\,.
\end{equation}
That is, isotropic $N$-flation is incapable of producing a parametrically super-Planckian effective decay constant for any choice of instanton action.

\subsection{Random Matrix N-flation} \label{sec:rmt}

In the previous section, we saw that the LWGC strongly constrains isotropic $N$-flation. These constraints are severe because in isotropic $N$-flation the axion fundamental domain is a hypercube, whose volume is famously much larger than that of an inscribed hypersphere.  A hypercube's side-length must be much smaller than a hypersphere's radius for the two to have comparable volume. It is thus natural to ask if we can achieve parametrically super-Planckian displacements, consistent with the LWGC, by changing the shape of the axion fundamental domain, or alternatively by changing the kinetic matrix $K_{ij}$ in the lattice basis.  In this section, we consider what happens when the kinetic matrix  is drawn from a random matrix ensemble. Since this matrix must be positive definite, a natural possibility is to take $K_{ij}$ to be a Wishart matrix, or alternatively an inverse-Wishart matrix (i.e., $K_{ij}^{-1}$ is a Wishart matrix).

\vskip 3pt
We begin by reviewing the definition and relevant properties of a Wishart matrix. Consider an $N \times M$ matrix $A$ whose entries are normally distributed with mean 0 and variance $\sigma^2$,
\begin{equation}
 A_{ij}\sim \mathcal{N}(0,\sigma^2)\,.
\end{equation}
Define the $N \times N$ matrix $K$ by
\begin{equation}
K= A A^\top.
\end{equation}
$K$ is then said to be an $N \times N$ Wishart matrix with $M$ degrees of freedom, while $K^{-1}$ is then said to be an inverse-Wishart matrix.
\vskip 3pt
Consider an ensemble $\cal{E}$ of $N\times N$ matrices.  Suppose that ${\cal E}$  is statistically rotationally invariant, so that the corresponding normalized eigenvectors $\psi_a$ point in directions  that are uniformly  distributed on $S^{N-1}$.
Then, in the large $N$ limit, the components $\psi_{a}^{(i)}$, $i=1,\ldots,N$,
are distributed as
$\sqrt{N} \psi_{a}^{(i)} \in {\cal N}(0,1)$, with ${\cal N}(0,1)$ denoting the normal distribution with mean $0$ and variance $1$.
Intuitively,  a single component of order unity is possible only if many other components are atypically small.
More geometrically, nearly all eigenvectors point approximately along a diagonal direction in some hyperoctant,  rather than being nearly parallel to a Cartesian basis vector.  This is not surprising, since there are $2^N$ diagonals but just $N$  basis vectors.
This phenomenon is known as {\it{eigenvector delocalization}}  in random matrix theory, and has been proved to hold in a number of random matrix ensembles \cite{Erdos09,TaoVu}, including the Wishart ensemble. It was argued in \S4 of \cite{Bachlechner:2014gfa} that this phenomenon may play an important role in generating large diameters in string theory.\footnote{Related ideas are examined in \cite{Bachlechner:2017hsj}, by means of new methods for analyzing many-axion landscapes.}

\vskip 3pt
Returning to the axion model in (\ref{eq:latticebasis}), let us work in the lattice basis and suppose that the kinetic matrix $K_{ij}$ is drawn from a Wishart ensemble; that is, $K_{ij}$ is a Wishart matrix (we leave the variance unspecified for now). Let $f_i^2$ be the $i$-th eigenvalue,
with $f_1 < f_2 < ... < f_N$.
The fundamental domain of axion field space is an $N$-cube in the lattice basis. Eigenvector delocalization tells us that the eigenvector $\psi_N$ with eigenvalue $f_N^2$  points along a nearly diagonal direction \cite{TaoVu}, along which ${\cal F}$  has diameter $2\pi\sqrt{N}$.  As a result, the diameter of the fundamental domain is
\begin{equation}
{\mathcal{D}} \equiv 2 \pi f_{\text{eff}} \approx 2\pi f_N \sqrt{N} \, ,  \label{diam2}
\end{equation}
where the $\approx$  becomes an equality in the case of perfect alignment. An important caveat is that we have estimated the diameter of the field space, but it is not always the case that there is an approximately flat direction of the potential nearly parallel to the long diameter of the field space. 
%\tru{Check my usage of expectation values $\langle ... \rangle$ below.}

\vskip 3pt
For a Wishart kinetic matrix $K_{ij}$, we have~\cite{Johnstone:2001wish,Goodman:1963ddc}
 \begin{equation}
\exp\,\Bigl\langle \log\left( \prod f_i^2\right) \Bigr\rangle = \left(\frac{ \langle f_N \rangle^2 }{4N}\right)^N \!\!\Gamma(N+1)\,,
\end{equation}
where $\langle ... \rangle$ indicates an expectation value in the Wishart ensemble.
Using Stirling's approximation we find
\begin{equation}
\log\Bigl(\lab{vol}\,\bm{\phi}/\M^N\Bigr) = N\log\left(\frac{\pi}{\sqrt{e}} \langle f_N \rangle \right)+{\cal{O}}(\log N)\,,
\end{equation}
so the volume bound (\ref{eq:volBound}) reads
\begin{equation}
\langle f_N \rangle \lesssim 2^{3/2} e\mu\pi^{-1/2}N^{-1/2}\,,
\end{equation}
so that parametrically,
\begin{equation}
\langle f_N \rangle \lesssim {\cal{O}}(N^{-1/2})\,.
\end{equation}
On the other hand, from (\ref{diam2}) we have the firm limit ${\mathcal{D}} \le 2\pi f_N \sqrt{N}$, and so
\begin{equation} \label{kwishvolb}
 {\mathcal{D}} \lesssim {\cal{O}}(1)\,,
\end{equation} when $K_{ij}$ is a Wishart matrix.

\vskip 3pt
Now, suppose instead that the kinetic matrix $K_{ij}$ is an inverse-Wishart matrix.
We define the aspect ratio ${\cal A}$ of the kinetic matrix by
\begin{equation}
{\cal A} = \frac{f_N}{({\rm{det}} K)^{1/{2N}}} =  \frac{f_N}{{( \prod f_i )}^{1/N}}\,.
\end{equation}
The point of considering this quantity is that for two different matrices (or matrix ensembles) $J$, $K$, the maximal diameters allowed by the volume bound have the ratio
\begin{equation}
\frac{{\mathcal{D}}_J}{{\mathcal{D}}_K} = \frac{{\cal A}_J}{{\cal A}_K}\,.
\end{equation}
We will now show that
\begin{equation}
{\cal A}({\rm{inverse~Wishart}}) \approx N {\cal A}({\rm{Wishart}})\,.
\end{equation}
For $J$ a Wishart matrix with associated variance $\sigma^2$, we have
\begin{equation}
\langle f_N^2 \rangle_{\rm{W}} = 4N\sigma^2\,.
\end{equation}
On the other hand, we argued above that
\begin{equation}
\Bigl\langle \prod f_i\Bigr\rangle^{1/N}_{\rm{W}} \approx \frac{\langle f_N \rangle_{\rm{W}}}{2\sqrt{e}}\,,
\end{equation} so that
\begin{equation}
{\cal A}({\rm{Wishart}}) \approx 2\sqrt{e} \sim {\cal{O}}(1)\,.
\end{equation}
The corresponding inverse Wishart matrix ensemble has
\begin{equation}
\langle f_N^2 \rangle_{\rm{IW}} = \frac{N}{c\sigma^2}\,,
\end{equation} with $c \approx 0.30$.
Thus,
\begin{equation}
{\cal A}({\rm{inverse~Wishart}}) = \frac{\langle f_N \rangle_{\rm{IW}}}{\langle \prod f_i\rangle^{1/N}_{\rm{IW}}} \approx \langle f_N \rangle_{\rm{IW}} \times \Bigl\langle \prod f_i\Bigr\rangle^{1/N}_{\rm{W}} = \sqrt{\frac{N}{c\sigma^2}} \times \frac{1}{2\sqrt{e}} \sqrt{4N\sigma^2}\,.
\end{equation}
In this approximation,
\begin{equation}
{\cal A}({\rm{inverse~Wishart}}) \approx \frac{N}{\sqrt{c e}}\,.
\end{equation}
Hence, when $K$ is an inverse Wishart matrix, the volume bound (\ref{eq:volBound}) translates to
\begin{equation} \label{kinvwishvolb}
 {\mathcal{D}} \lesssim {\cal{O}}(N)\,,
\end{equation}
so parametric enhancement of the diameter is compatible with the volume bound.

\vskip 3pt
The reason this enhancement can occur in the case of an inverse-Wishart but not a Wishart kinetic matrix is that the inverse-Wishart eigenvalue distribution has a heavier tail than the Wishart eigenvalue distribution: there is a significant probability for one or more eigenvalues of an inverse-Wishart matrix to be much larger or much smaller than the average. As a result, the fundamental domain will be squashed in some directions and stretched in others, thereby allowing the diameter in some directions to be very large while the total volume is small. For a Wishart kinetic matrix, on the other hand, the fundamental domain is shaped more like a cube, so no side length can be parametrically large unless the volume is too.

\vskip 3pt
Once again, we stress that the volume bound we have employed here is a necessary but not sufficient condition for a theory to satisfy the LWGC. Indeed, we derived the volume bound in the continuum approximation, but if the volume bound is obeyed, and if one eigenvalue of the kinetic matrix is parametrically super-Planckian, then the product of the other eigenvalues must be parametrically sub-Planckian. In this limit, the continuum approximation, and therefore the volume bound itself, becomes suspect, and technology beyond the scope of the present paper is needed. We conclude from our analysis that parametrically super-Planckian diameters for Wishart kinetic matrices are inconsistent with the LWGC, while the prospects for inverse-Wishart kinetic matrices are yet unclear. Nonetheless, by moving beyond the continuum approximation, we will show in the following section that there exist models of \emph{stretched N-flation} in which one eigenvalue of the kinetic matrix is taken to be parametrically larger than the others in a manner that is consistent with the LWGC.

\section{Inflation Compatible with the LWGC}\label{sec:Compatible}

In the previous section, we found that some simple models of axion inflation are highly constrained by the LWGC, and in particular by the volume bound~\eqref{eq:volConstraint}. However, per our discussion in \S\ref{sec:rmt}, the volume bound does not completely rule out parametrically super-Planckian field displacements. In this section, we show how loopholes in the above constraints allow models of axion inflation---including KNP alignment, clockwork, and modified $N$-flation---to evade these bounds. Many of these models require rather special structures in the effective theory,
%\lm{rephrased here}
and it is not obvious that they can be realized in an ultraviolet-complete framework like string theory. Still, the fact that they exist as effective theories demonstrates that the LWGC alone is insufficient to rule out models of axion inflation with super-Planckian field displacements, even in the absence of monodromy.

\subsection{Coherent Instanton Sums}\label{ssec:coherent}

In the previous section, we assumed that the instanton phases were randomly distributed, so that $\langle Z_\mb{Q}\rangle = 0$.  The first loophole in the LWGC constraints occurs when the phases are not random, but are fixed so that the contributions to the potential sum in a coherent way.

\vskip 3pt
We begin with a single-axion theory with charge lattice $\Gamma = \{\, n/f\,| \,n \in \mathbbm{Z}\}$, whose potential in the canonically normalized basis is
\begin{equation}
    V(\phi) = \sum_{n} Z_n e^{i n \phi/f}\,,
\end{equation}
where we take
\begin{equation}
    Z_{n} = Z_{\sminus n}^* = V_1 e^{-\alpha n + i \delta_n}\, , \mathrlap{\qquad n > 0\,,}
\end{equation}
and $Z_0 = V_0$. The Lagrangian is then
\begin{equation}
\mathcal{L} = -\frac{1}{2} (\partial \phi)^2 -V_0- 2 V_1 \sum_{n=1}^\infty e^{-\alpha n} \cos\bigl( n \phi/f+ \delta_n \bigr)\,.
\label{eq:coherentsum}
\end{equation}

\begin{figure}
\begin{center}
\hspace{-1cm}\includegraphics{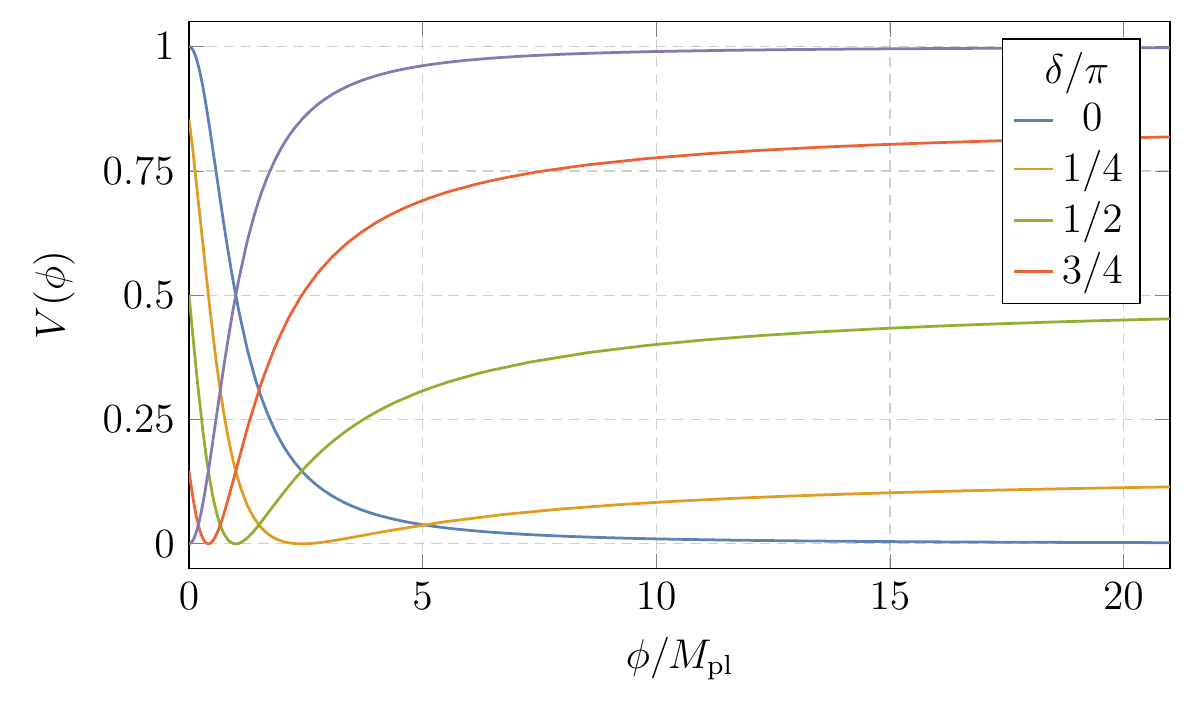}
\caption{Coherent sum potentials with $\alpha \ll 1$ for various values of $\delta$. A sum over instantons with equal phases produces a smooth potential capable of supporting large-field inflation. 
%\tru{I changed the second sentence of this caption: I think the the old one: ``For a wide range of values, $\pi/8 \lesssim \delta \lesssim 9\pi/8$, the model predicts a tensor-to-scalar ratio and a spectral index in good agreement with current observational constraints.'' was too similar to the caption of figure 4, and didn't really describe this figure.}
\label{fig:coherentsum}}
\end{center}
\end{figure}

\vskip 3pt
In a single-axion theory, the LWGC implies that $\alpha \gtrsim M_\lab{pl}/f$ so that, for $f \gg 1$, $\mathcal{O}(f/ M_{\lab{pl}})$ terms in the sum contribute appreciably to the potential. If the phases $\delta_n$ are independently and randomly distributed, incoherent addition of $\mathcal{O}(f/ M_{\lab{pl}})$  summands will generically spoil the flatness of the potential on super-Planckian distances, and large-field inflation will be impossible. However, the story is very different if we take all of the phases to be equal, $\delta_n \equiv \delta$. In this case, we can explicitly compute the sum in (\ref{eq:coherentsum}),
\begin{equation}
V(\phi) = V_0 + V_1 \left(\frac{\cos(\phi/f+ \delta)-e^{-\alpha} \cos \delta}{\cosh \alpha - \cos \phi/f}\right).
\end{equation}
Note that the transformation $\delta \rightarrow \delta + \pi$ is equivalent to $V_1 \rightarrow - V_1$, so we may assume that $V_1 > 0$ without loss of generality. Similarly, the transformation $\delta \rightarrow - \delta$ is equivalent to $\phi \rightarrow - \phi$, so we may without loss of generality focus on the part of the potential with $V'(\phi) > 0$, so that $\phi$ decreases over the course of its slow-roll. We further suppose that $V_0$ is chosen so that the the potential vanishes at its minimum. For concreteness, we will set $f = M_\lab{pl}/\alpha$. Now, we have an entire family of potentials parametrized by $V_1$, $\delta$, and $\alpha$. For $\alpha \gg 1$, the first term in the sum in (\ref{eq:coherentsum}) dominates, and as expected we get a simple cosine potential with sub-Planckian decay constant. More interesting is the $\alpha \ll 1$ limit, which the LWGC typically constrains. However, in this limit the potential, to leading order in $\alpha$, is
\begin{equation}
  V(\phi) = \frac{2 V_1}{\alpha} \frac{(M_\lab{pl} \cos \delta/2 - \phi \sin \delta/2)^2}{M_\lab{pl}^2 + \phi^2}\,.
\end{equation}
In this limit, $\alpha$ simply contributes to the overall scaling of the potential and can therefore be absorbed into the overall normalization $V_1$. If we assume this normalization is fixed by the observed value of the power spectrum, we are left with just a single-parameter family of potentials, labeled by $\delta$. This potential is shown in Figure~\ref{fig:coherentsum} for several values of $\delta$.

\vskip 3pt
We can compute the phenomenology of these models in the slow-roll approximation. The slow-roll parameters $\epsilon$ and $\eta$ are defined by
		\begin{equation}
			\epsilon \equiv \frac{M_\lab{pl}^2}{2} \left(\frac{V'}{V}\right)^2 \qquad \eta \equiv M_\lab{pl}^2 \frac{V''}{V}\,,
		\end{equation}
		and the number of $e$-folds is given by
		\begin{equation}
			N_* = \int_{\phi_\lab{end}}^{\phi_*} \frac{\ud \phi}{M_\lab{pl}} \frac{1}{\sqrt{2 \epsilon}}\,. 					\end{equation}
		We can then compute the spectral index and tensor-to-scalar ratio to first order in the slow-roll approximation,
		\begin{align}
		n_s = 1 + 2 \eta_* - 6 \epsilon_* \,, \quad r = 16 \epsilon_*\,,
		\end{align}
where the $*$ indicates that we are evaluating at horizon crossing, roughly 50-60 $e$-folds before the end of inflation.

\begin{figure}
\begin{center}
\includegraphics{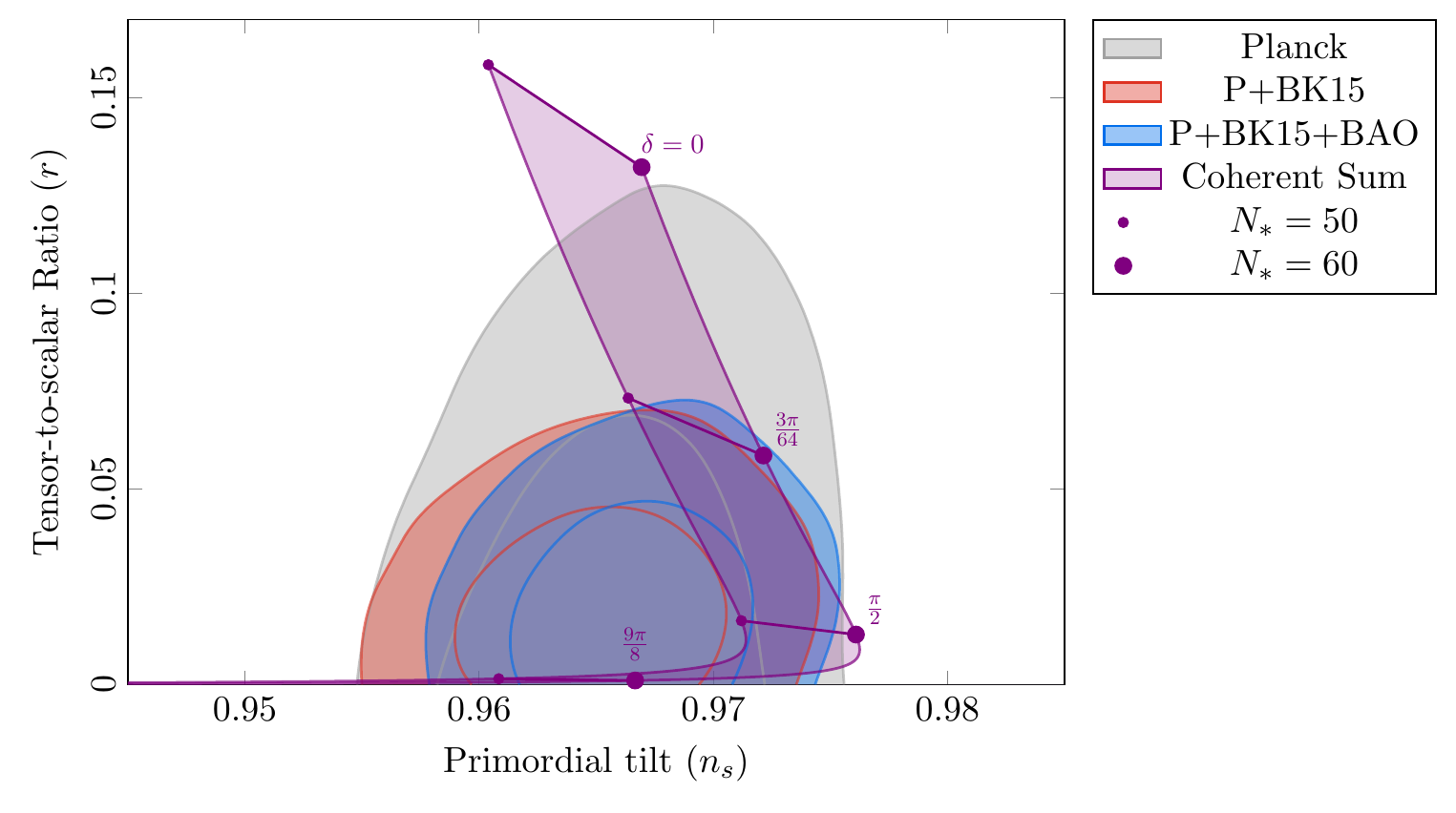}
\caption{The predicted values of the spectral index $n_s$ and tensor-to-scalar ratio $r$ for various values of $\delta$ at a pivot scale 50 $e$-folds and 60 $e$-folds before the end of inflation, compared with the \emph{Planck} $1\sigma$ and  $2\sigma$ exclusion limits. For a wide range of values $3\pi/64 \lesssim \delta  \lesssim 9\pi/8$, the predicted values agree well with the data.} \label{fig:coherentsumpheno}
\end{center}
\end{figure}

\vskip 3pt
The phenomenology of this family of models is in comfortable agreement with constraints from measurements of the CMB, as shown in Figure~\ref{fig:coherentsumpheno}. For $\delta$ between $3\pi/64$ and $9 \pi /8$, the spectral index (measured at a pivot scale 50 $e$-folds before the end of inflation) lies between $n_s \approx 0.96$ and $n_s \approx 0.98$, and the tensor-to-scalar ratio lies between $r=0.001$ and $r=0.010$. These values are compatible with the constraints determined by \emph{Planck}, which found $n_s = 0.9649 \pm 0.0042$ (68\% CL) and $r < 0.10$ (95\% CL) \cite{Akrami:2018odb}.

\vskip 3pt
One might worry that the precise notion of an instanton, and hence the WGC bound we are invoking, breaks down in the $\alpha < 1$ limit we considered here (see footnote \ref{footinst} in \S\ref{sec:AxionsLattices}). However, as discussed in \S\ref{ssec:sLWGC}, we could imagine that these instantons descend from five-dimensional charged particles, in which case the WGC bound---and the instanton sum---makes sense even for small values of the action. Indeed, in such a scenario, the phases $\delta_n$ are given simply by $(-1)^F$ \cite{hosotani:1983xw}, so if the tower of charged particles are all light bosons (or fermions), they will give rise to the same coherent sum behavior we have seen here.

\vskip 3pt
One can also generalize this single-axion model to a theory with multiple axions, including an isotropic $N$-flation model.
In this case, the effective decay constant in the diagonal direction is enhanced by a factor of $\sqrt{N}$ relative to the decay constant $f$ in the basis directions, so the parameter $\alpha \sim \M / f$ and the effective decay constant $f_{\rm eff} = \sqrt{N} f$ can be large simultaneously.

\subsection{KNP Alignment}

The second loophole relies on using multiple ``species'' of instantons, some of which provide a much larger contribution to the inflationary potential than others. This can occur if, for instance, these special instantons have a much larger prefactor $A_{\bm{\ell}}$ or a much smaller action $S_{\bm{\ell}}$ than others. In what follows, we consider the latter possibility and use this ``multiple species loophole'' to realize the two-axion alignment mechanism of Kim, Nilles, and Peloso (KNP)~\cite{kim:2004rp}. In particular, we will consider the model described in~\cite{Delafuente:2014aca}.

\vskip 3pt
Let us first describe why this type of alignment can enhance the effective field range, as the idea will also be useful for the following section. Throughout this paper, we have assumed that the axion potential is generated by summing over a fully populated (sub)lattice of instantons---this is the natural situation for a set of periodic scalars. However, it may be the case that, for some physical reason, this sum truncates so that only $N$ charges $\mb{Q}_{k}$ contribute, where $k = 1, \dots, N$.  Then, the canonically-normalized Lagrangian takes the form
\begin{equation}
    \mathcal{L} = -\frac{1}{2} \delta_{ab}\, \partial_\mu \phi^a \, \partial^\mu \phi^b - \Lambda^4\sum_{k=1}^{N} \bigl(1 - \cos(Q_{k, a} \phi^a/f)\bigr)\,. \label{eq:truncLag}
\end{equation}
Here we have, for simplicity, assumed that these terms share the same prefactors $Z_{\mb{Q}}$, with phases such that the potential is minimized at $\phi^i = 0$. We can quickly characterize the maximum effective field range in (\ref{eq:truncLag}) by studying the eigenvalues of the axionic mass matrix,
\begin{equation}
    M_{ab}^2 \equiv \frac{\Lambda^4}{f^2} \sum_{k = 1}^{N} Q_{k, a} Q_{k, b}\,.
\end{equation}
We may define the effective axion decay constants in terms of the eigenvalues of this mass matrix,
\begin{equation}
  f_{\lab{eff}, k}^{2} = \Lambda^{4}\,\,\lab{eig}_k\, M^{-2}_{ab}\,, \label{eq:effectiveDecayConstants}
\end{equation}
as we may also think of this quadratic approximation as descending from the potential
\begin{equation}
  V(\phi^i) = \Lambda^4 \sum_{k = 1}^{N} \left(1 - \cos(\varphi^k/f_{\lab{eff},k})\right)\,,
\end{equation}
where the $\varphi^k$ represent linear combinations of the $\phi^a$ corresponding to eigenvectors of $M_{ab}$.
We must stress that the definition of the effective axion decay constants is basis-dependent and ceases to make sense once we deviate from the specific form of (\ref{eq:truncLag}), though it will be helpful in what follows.

\vskip 3pt

\begin{figure}
\centering
\includegraphics{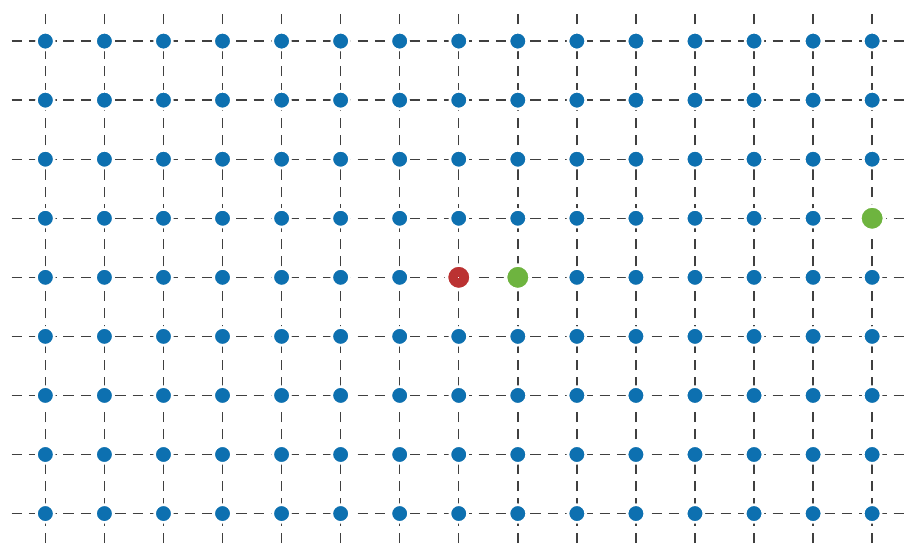}
\caption{KNP alignment consistent with the LWGC. Here, the blue dots indicate charge sites with marginally superextremal instantons, the red dot indicates the origin $\mb{Q} = 0$, and the large green dots of charge $(1,0)$, $(N,1)$ indicate charge sites with very superextremal instantons, whose potential contributions dominate the inflationary potential and produce a large effective decay constant.}
\label{fig:KNP}
\end{figure}

The KNP alignment mechanism\footnote{KNP alignment has also been called lattice alignment to distinguish it from kinetic alignment, as in \cite{bachlechner:2014hsa}. In kinetic alignment, the non-trivial structure appears in the kinetic matrix instead of the charge lattice. Of course, these two cases are related to one another via a basis transformation.} relies on a clever choice of charges $Q_{k,a}$ to attain a very large effective axion decay constant. Let us consider the example presented in~\cite{Delafuente:2014aca} and pictured in Figure~\ref{fig:KNP}. Here, two instantons with charges $(1,0)$ and $(N, 1)$ generate the potential. Because these contributions are very nearly parallel, the $(1, -N)$ direction in field space will be much lighter, and have much larger effective axion decay constant $f_{\lab{eff}, 1} \sim N f$,  than the orthogonal direction, with $f_{\lab{eff}, 2} \sim f/N$.

\vskip 3pt

From Figure~\ref{fig:KNP}, we can see how the LWGC constrains KNP alignment: the success of the model depends crucially on the truncation in (\ref{eq:truncLag}), which allows us to ignore lattice sites that are much closer to the origin---and thus potentially much more important---than the instanton with charge $(N, 1)$. Once we include the complete lattice sum, any alignment will be destroyed, unless we make the ``aligning'' species of instanton much stronger than the rest of the lattice sum.

\vskip 3pt
It will be useful to understand how a background lattice of LWGC-fulfilling instantons, pulled from the ensemble discussed in \S\ref{sec:volume}, can affect this type of extra-species KNP alignment. The extremal instantons contribute to the harmonic variance
\begin{equation}
  \sigma_n^2 = \sum_{\ell_2 \in \mathbbm{Z}} e^{- 2 \alpha (M_\lab{pl}/f) \sqrt{(n + \ell_2 N)^2 + \ell_2^2}}\,, \mathrlap{\qquad n > 1\,,}
\end{equation}
where we have taken $\mu = \alpha M_\lab{pl}$ and $\bar{\mb{e}} = f(1, \minus N)$ in (\ref{eq:harmVariance}).
The most significant harmonics are $n = 1$, which gets its main contribution from the $\ell_2 = 0$ term
\begin{equation}
  \sigma_1^2 \simeq e^{-2 \alpha M_\lab{pl}/f}\,,
\end{equation}
and $n = k N$, whose main contribution comes from instantons with charge $(0, \minus k)/f$,
\begin{equation}
    \sigma_{k N}^2 \simeq e^{-2 \alpha k M_\lab{pl}/f}\,.
\end{equation}
These harmonics are sub-Planckian in scale for $f \lesssim M_{\text{pl}}$, and
can be controlled by taking $f / M_{\text{pl}} \ll 1$. For instance, if we
demand a power law suppression $\sigma_n^2 \lesssim \frac{1}{n^p}$, we must have that
\begin{equation}
  f \lesssim \frac{2}{p \log N} M_{\text{pl}}\,,
\end{equation}
which in turn implies that the effective axion decay constant is bounded by
\begin{equation}
  f_{\text{eff}} \lesssim \frac{2 N}{p \log N} M_{\text{pl}}\,. \label{eq:knpFeff}
\end{equation}
Thus, even after including a lattice of LWGC-fulfilling instantons, the effective field range is still super-Planckian for $N \gg 1$.

\vskip 3pt
We conclude that the KNP alignment mechanism is consistent with the LWGC. However, it is
clear that if the instanton factors $Z_{\mb{Q}}$ are drawn from a random
distribution with ${\langle | Z_{\mb{Q}} |^2 \rangle \sim e^{- 2 \alpha M_\lab{pl} | \mb{Q} |}}$ as in \S\ref{sec:volume}, rather than by postulating light
instantons with charges $(1, 0)/f$ and $(N, 1)/f$ by hand, then the likelihood of
getting a suitably aligned potential is extremely low. Furthermore, it is unclear if there exists a mechanism to enhance certain instanton contributions over others, especially in a theory coupled to gravity. Thus, although this model is not incompatible with the LWGC, finding an ultraviolet completion seems quite non-trivial and so it may yet reside in the Swampland.

\subsection{Clockwork Alignment}

As discussed in the previous section, KNP alignment relies on a special structure in the truncated charge matrix $Q_{k,a}$, cf.~(\ref{eq:truncLag}), to achieve an enhanced effective decay constant. The \emph{clockwork} mechanism, first proposed by \cite{Choi:2014rja,Choi:2015fiu,Kaplan:2015fuy}, can be thought of as KNP alignment iterated to achieve vastly super-Planckian effective axion decay constants from multiple sub-Planckian axions. In this section, we will first review what makes this mechanism tick and then describe how the LWGC constrains it.

  \vskip 3pt
  We will focus on the construction presented in~\cite{Kaplan:2015fuy}, which achieves the charge matrix
  \begin{equation}
    Q^\lab{CW}_{k, a} =
    \begin{pmatrix}
      \minus 1 & 3 & 0 & 0 & \cdots & 0 & 0 \\
      0 & \minus 1 & 3 & 0 & \cdots & 0 & 0 \\
      0 & 0 & \minus 1 & 3 & \cdots & 0 & 0 \\
      \vdots & \vdots & \vdots & \vdots & \ddots & \vdots & \vdots \\
      0 & 0 & 0 & 0 & \cdots & \minus 1 & \es\,3\es\, \\
      0 & 0 & 0 & 0 & \cdots & 0 &  c
    \end{pmatrix}, \label{eq:clockwork}
  \end{equation}
  via a specially crafted pattern of spontaneous symmetry breaking, though their specific ultraviolet completion will not be relevant to our story.\footnote{See also Appendix \ref{sec:StringTheory} for comments on the prospects for finding such a completion in string theory.}
  They find that the largest effective decay constant (\ref{eq:effectiveDecayConstants}) is \emph{exponentially} enhanced in the number of axions $N$,
  %\js{scaled $c$}
  \begin{equation}\label{eq:bigeigc}
    f_\lab{eff}^2 = f^2 \left(1 + \frac{8}{c^2}\right)\frac{3^{2 N} - 1}{64} \,,
  \end{equation}
  and they give an elegant explanation for this anomalously large eigenvalue based on an emergent translational symmetry. However, it will be useful to work with an alternative explanation that will make it obvious that the clockwork mechanism cannot survive the LWGC unless it relies on the previous section's ``multiple species'' loophole.

  \vskip 3pt
  Where does the anomalously large eigenvalue \eqref{eq:bigeigc} come from? We can answer this by first noting that the determinant
  \begin{equation}
    \left|\lab{det}\, Q_{k,a}^{\lab{CW}} \right| =  |c|\,,\label{eq:detMat}
  \end{equation}
  is both independent of $N$ and becomes unity when $c = \pm 1$. For these special values of $c$, this charge matrix is a member of $\lab{GL}(N, \mathbbm{Z})$.  For general $c$, (\ref{eq:clockwork}) thus parameterizes a family of matrices ``near'' this element, in the sense that small changes in $c$ still keep the determinant (\ref{eq:detMat}) close to one.
  \vskip 3pt
  Elements of $\lab{GL}(N, \mathbbm{Z})$ are special in that their determinants, $\pm 1$, are anomalously small.  For example, if the charge matrix is instead a random $N \times N$ Bernoulli matrix, whose entries are $\pm 1$, the log of the absolute value of its determinant is normally distributed about $\log \sqrt{(N-1)!}$ with variance $\log \sqrt{N}$ \cite{Nguyen:2011rml}. Clearly, large integer matrices with unit determinant are very rare!

  \vskip 3pt
  Since the mass matrix is essentially the square of the charge matrix $Q_{k,a}^{\lab{CW}}$, we may attribute its anomalously small eigenvalue to its similarly small determinant---since all other eigenvalues range from $(3-1)^2$ to $(3+1)^2$, the smallest must be exponentially small to ensure the determinant is close to unity. It is easy to generate examples of charge matrices $Q_{k,a}$ that realize clockwork-style alignment---take one's favorite element of $\lab{GL}(N, \mathbbm{Z})$ and perturb it in a way that does not appreciably disturb the determinant.\footnote{It is not always the case that an element of $\lab{GL}(N, \mathbbm{Z})$ has a single anomalously small eigenvalue, as we can simply consider the inverse of (\ref{eq:clockwork}) which has many small eigenvalues and a single enormous one. It is thus more appropriate to say that these matrices typically have an enormous hierarchy between one eigenvalue and the rest.}

  \vskip 3pt
  Unfortunately, the clockwork mechanism clearly fails once we include a lattice of charges.\footnote{One might hope that the field theory realization presented in \cite{Kaplan:2015fuy} avoids these constraints, as it does not rely on instantons to construct its potential. However, as we explain in Appendix \ref{sec:StringTheory}, ultraviolet realizations of this model that rely on extra-dimensional locality are susceptible to similar problems---there are important desequestering effects that must be suppressed
%\lm{rephrased}
  in order to realize (\ref{eq:clockwork}).}  The fact that the charge matrix is very nearly an element of $\lab{GL}(N, \mathbbm{Z})$ means that we have expressed the charge lattice in an extremely misleading basis---there are lattice sites that are much closer to the origin, and thus provide large corrections to the Lagrangian, that we have ignored in our truncation to (\ref{eq:truncLag}). As in Figure~\ref{fig:KNP}, these ignored-but-dominant contributions will drastically reduce the effective field range over which we can inflate.

  \vskip 3pt
  We can again circumvent this problem if we assume that the lattice sites generating the clockwork alignment are members of a different, dominant species. Constraints similar to (\ref{eq:knpFeff}) then follow if we assume that the LWGC is instead satisfied by a background lattice of instantons, which again allows for super-Planckian effective field ranges. However, we should again emphasize that we lack a compelling mechanism to enhance particular lattice sites over others. Furthermore, even with such a mechanism, one would need to explain why this type of iterated alignment---which comes at an enormous statistical price---would be generated by the enhanced species.

\subsection{Stretched N-flation} \label{sec:stretched}

In \S\ref{ssec:isotropic}, we argued that isotropic $N$-flation cannot achieve parametrically super-Planckian displacements while simultaneously satisfying the volume bound \eqref{eq:volBound} required by the LWGC. However, in \S\ref{sec:rmt} we found that the volume bound could be circumvented by skewing the axion fundamental domain. Non-hypercubic fundamental domains can simultaneously obey the volume bound and contain a direction that is parametrically super-Planckian. That analysis came with a caveat, however, as the volume bound is only meaningful when the continuum approximation applies, and there may be directions that should not be Poisson resummed if the fundamental domain is extremely skewed. Properly analyzing these cases requires technology \cite{Resummation} beyond the scope of this work. In this section, we will go beyond the continuum approximation for a simple variation of $N$-flation, which we term \emph{stretched $N$-flation}, using the techniques of \S\ref{ssec:isotropic}, and we will show that this model indeed produces parametrically super-Planckian directions consistent with the LWGC.

\begin{figure}
\centering
\includegraphics{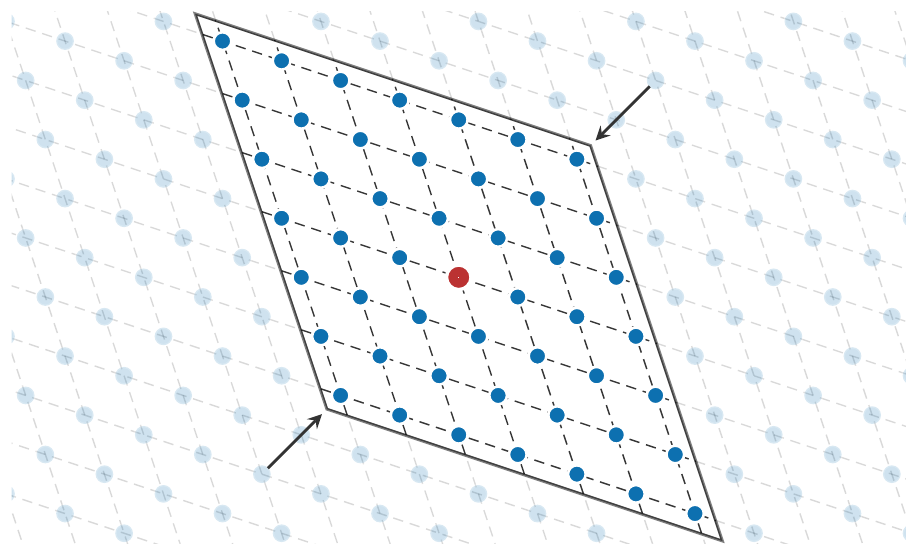}
\caption{The charge lattice for stretched $N$-flation. One direction in field space is stretched relative to the others, leading to a charge lattice that is slightly squashed in one direction. At large $N$, this allows a parametrically-enhanced effective decay constant in the stretched direction while satisfying the LWGC.}
\label{fig:stretched}
\end{figure}

\vskip 3pt
As the name suggests, the general idea is to stretch isotropic $N$-flation to accommodate both the volume bound and a parametrically super-Planckian direction. Isotropic $N$-flation's charge lattice is generated by
\begin{equation}
    \mb{f}^{\es\sminus 1} = \mathbbm{1}/f\,,
\end{equation}
so that $\mb{Q} = \mb{f}^{\es\sminus 1} \bm{\ell} = (f^{\es \sminus1})\indices{^i_a} \ell_i$ is in the charge lattice $\Gamma$ for all $\bm{\ell} \in \mathbbm{Z}^N$, cf. \S\ref{sec:AxionsLattices}. The stretched charge lattice is instead generated by
\begin{equation}\label{eq:strlat}
  \mb{f}^{\es\sminus 1} = f_{2}^{\es \sminus 1} \mathbbm{1} + \frac{1}{N}\left(\frac{1}{f_1} - \frac{1}{f_2}\right) \mb{d}^\top \hspace{-0.5pt}\mb{d}\,,
\end{equation}
where $\mb{d} = (1,1, \dots, 1)$ is the isotropic charge lattice's minimal length lattice vector in the diagonal direction. Clearly, the stretched lattice \eqref{eq:strlat} reduces to the isotropic charge lattice when $f_1 = f_2 = f$. However, when $f_1 \neq f_2$, the lattice is stretched along the $\bar{\mb{e}} = \mb{f}^{\es\sminus 1} \mb{d}$ direction, so that the new diagonal has length
\begin{equation}
  |\bar{\mb{e}}| = \sqrt{N} f_1 = \sqrt{N} \beta f_2 \,,
  \label{eq:newLength}
\end{equation}
while the volume of the unit cell,
\begin{equation}
    |\Gamma| = \frac{f_2}{f_1} \frac{1}{f_2^{N}} = \beta \left(\frac{\alpha}{\mu}\right)^{N} , \label{eq:newUnitCell}
\end{equation}
is more sensitive to $f_2$ than to $f_1$. Here we have defined the dimensionless parameters $\alpha \equiv \mu/f_2 \simeq M_\lab{pl}/f_2$ and $\beta \equiv f_2/f_1$. The new, stretched charge lattice is schematically depicted in Figure~\ref{fig:stretched}.

\vskip 3pt

Using (\ref{eq:newLength}) and (\ref{eq:newUnitCell}), the continuum limit harmonic variances (\ref{eq:prHarmVar}) are
\begin{equation}
  \sigma_n^2 \simeq \frac{2 \alpha}{\pi \sqrt{N}}  \left( \frac{\beta \pi n}{\alpha \sqrt{N}} \right)^{\frac{N}{2}}\!\! K_{\frac{N}{2}} \left( \frac{2  \alpha \beta n}{\sqrt{N}} \right)\,, \label{eq:harmVarStretched}
\end{equation}
which simplify in the large-$N$ limit and are, crucially, independent of $\beta$,
 \begin{equation}
  \sigma_n^2 \simeq  \frac{\sqrt{\pi}}{\alpha e}  \left( \frac{\pi N}{2\alpha^2 e} \right)^{\frac{N - 2}{2}}, \qquad \alpha \beta n \ll \frac{N}{2 \sqrt{2}}\,.
 \end{equation}
We can thus heavily stretch the lattice along the diagonal by taking $\beta \ll 1$ while, at the same time, keeping the harmonic variances under control, with $\sigma_n^2 \ll 1$, by fixing $\alpha  \gg \sqrt{\pi N/(2 e)}$\, to satisfy the volume bound.
 If we define $\alpha = \hat{\alpha} \sqrt{N}$, we find that the length of the diagonal is $|\bar{\mb{e}}| \simeq  M_\lab{pl}/(\hat{\alpha} \beta)$ and can be vastly super-Planckian,
 \begin{equation}
    |\bar{\mb{e}}| \gg M_\lab{pl} \quad\text{when}\quad \beta \equiv \frac{f_2}{f_1} \ll 1 \quad \text{and}\quad \hat{\alpha} \simeq \frac{M_\lab{pl}}{\sqrt{N} f_2} \gg \sqrt{\frac{\pi}{2 e}}\,.
 \end{equation}
If $f_2$ is not sufficiently small, the volume of the fundamental domain will be too large and dangerous higher harmonics will reduce the field range.

\vskip 3pt
As alluded to in \S\ref{sec:rmt}, the continuum approximation can fail if there are some long directions in the charge lattice $\Gamma$. We should thus worry that (\ref{eq:harmVarStretched}) does not accurately capture the true magnitude of the LWGC-mandated higher harmonics, as the lattice is stretched more and more. However, as in \S\ref{ssec:isotropic}, we can also compute the harmonic variances by restricting the lattice sum to the instantons of smallest charge.\footnote{We could also use the richer saddle-point approximation developed in Appendix~\ref{sec:saddle}. However, this does not substantially change the result.} As we saw in Figure~\ref{fig:sigmavsk}, this provides information beyond the continuum limit and is a useful estimate of the terms in (\ref{eq:prHarmVar}) that we have dropped.

\vskip 3pt
Similar to~(\ref{eq:harmVarS}), the harmonic variances can be written as
\begin{equation}
  \sigma_n^2 =  \sum_s k^{(N, n)}_s e^{- 2 \alpha \sqrt{s + (\beta^2 - 1)n^2/N}}\,,
\end{equation}
where $k^{(N,n)}_n$ denotes the number of lattice sites $\mb{Q} = \mb{f}^{\es\sminus 1} \bm{\ell}$ that are a distance $\bm{\ell}^2 = s$ away from the origin in $\mathbbm{Z}^N$. The smallest charge $s = n$ instantons thus contribute
\begin{equation}
  \sigma_n^2 \sim \binom{N}{n} e^{- 2 \alpha \sqrt{n - (1 - \beta^2)  \frac{n^2}{N}}}\,,
  \qquad  (n \leq N)\,,
\end{equation}
and, in particular,
\begin{equation}
  \sigma_N^2 \sim e^{- 2 \alpha \beta \sqrt{N}}\,.
\end{equation}
To keep this harmonic suppressed relative to the leading $n = 1$ term, we must have that
\begin{equation}
  \beta \gtrsim 1 / \sqrt{N}\,,
\end{equation}
so that the field range is limited to
\begin{equation}
  |\bar{\mb{e}}| \simeq \frac{M_\lab{pl}}{\hat{\alpha} \beta} \lesssim \sqrt{N} M_\lab{pl}\,.
\end{equation}
Thus, unlike its isotropic sibling, stretched $N$-flation may realize a parametrically super-Planckian field range consistent with the LWGC.

\section{Conclusions} \label{sec:CONC}

We have studied axion potentials resulting from summation over a lattice of instantons.
This allowed us to analyze the constraints that the (s)LWGC imposes on axion inflation.

\vskip 3pt
We showed that if one requires that higher harmonics in the potential are suppressed, the LWGC implies a bound on the volume of the fundamental domain of axion field space.
This volume bound was invisible in analyses that approximated the axion potential by a sum over the instantons with the smallest charges: resummation was necessary to reveal the constraint.  Isotropic $N$-flation is incompatible with the volume bound, but we exhibited a new model, \emph{stretched} $N$-flation, that enjoys parametric enhancement of the axion field space while remaining consistent with the volume bound, and more generally with the LWGC.
We also showed that a coherent sum of instantons with no relative phases could produce a phenomenologically-viable potential.

\vskip 3pt

We also examined the impact of the LWGC on the KNP alignment mechanism and the clockwork mechanism.  We showed that these constructions can support large-field inflation while remaining compatible with the LWGC only in theories with at least two ``species'' of instantons: then the LWGC can be fulfilled by a lattice of instantons that make subleading contributions to the potential, while a few instantons of a different species create an aligned sector that dominates in the potential.  We pointed out that such structures appear statistically improbable in ensembles of effective theories, especially for the clockwork mechanism.
%\lm{rephrased}

\vskip 3pt

In summary, the LWGC---the strongest conjectured version of the WGC---imposes severe constraints on certain models of large-field axion inflation.   Nevertheless, the LWGC is \emph{not} sufficient to rule out all effective theories of large-field axion inflation, even those without monodromy.

\vskip 3pt

What is not clear so far is whether the models discussed here that evade the LWGC can actually be embedded in an ultraviolet-complete framework like string theory and, if so, how generic they are.  Confronting this problem in string theory is an important task for the future.

\section*{Acknowledgments}

We thank Thomas Bachlechner, Naomi Gendler, and Irene Valenzuela for helpful conversations, and we are indebted to Matt Reece for collaboration in the early stages of this work and for comments on a draft.  The work of B.H.\ was supported in part by NSF grant PHY-1914934 and in part by Perimeter Institute for Theoretical Physics. Research at Perimeter Institute is supported by the Government of Canada through Industry Canada and by the Province of Ontario through the Ministry of Economic Development and Innovation. The work of C.L.~was supported by NSF grant PHY-1848089. The work of L.M.~was supported in part by NSF grant PHY-1719877.  The work of T.R.~was supported by NSF grant PHY-1911298 and the Roger Dashen Membership. The work of J.S. is part of the Delta-ITP consortium, and is supported by a Vidi grant of the Netherlands Organisation for Scientific Research (NWO) that is funded by the Dutch Ministry of Education, Culture, and Science (OCW). Portions of this work were completed at the Aspen Center for Physics, which is supported by National Science Foundation grant PHY-1607611.

%\newpage
\appendix
\addtocontents{toc}{\protect\setcounter{tocdepth}{1}}

\section{Axion Inflation in String Theory}\label{sec:StringTheory}

In the main text, we have analyzed the impact of the WGC on inflation in effective theories of axions coupled to gravity.  Although much of the evidence for the WGC rests on properties of presently-known solutions of string theory, thus far we have made little use of such ultraviolet information.  We have instead treated the WGC as an infrared test, specifically as a (candidate) necessary condition for ultraviolet completion in quantum gravity, and we have imposed no other restrictions on the infrared theory.

\vskip 3pt
However, it is probable that string theory \emph{does} impose additional restrictions, so that the WGC alone is insufficient to ensure that an effective theory admits an ultraviolet completion in string theory.  To identify the classes of effective theories that actually arise in string theory, a practical approach is top-down construction, i.e., the enumeration of solutions of string theory.

\vskip 3pt
In this appendix we will describe the prospects for finding, in solutions of string theory, the effective theories of large-field axion inflation that were discussed in the preceding sections.
We should mention at the outset that certain broad properties of axion effective theories are strongly supported in string theory.
In particular, the existence of many axions with periodic scalar potentials generated by nonperturbative effects is a generic finding in, for example, compactifications of type II string theories on Calabi-Yau orientifolds.  The more difficult question is whether the patterns of couplings and characteristic scales that allow for large-field inflation arise in a computationally accessible regime.

\vskip 3pt
At the time of writing there is no universally acclaimed de Sitter solution of string theory, though there are scenarios that have withstood intense scrutiny, and within which many workers expect that solutions can be found. Inflationary solutions are more difficult to construct than de Sitter solutions, because of the nontrivial dependence on time.
Our remarks here should therefore be understood as characterizing properties of the effective theories that can be derived from compactifications of string theory, rather than properties of a (not yet constructed) ensemble of totally explicit inflationary solutions.

\subsection{Enumeration of axion theories}

We will focus our comments on axions arising from the Ramond-Ramond four-form $C_4$ in flux compactifications of type IIB string theory on O3/O7 orientifolds of Calabi-Yau threefolds, in the regime of weak coupling and large volume, because moduli stabilization is comparatively well understood in this context.
In principle, one could aim to enumerate axion effective theories in this corner of string theory as follows.  First, identify a tractable category of compactification manifolds, such as orientifolds of complete intersection Calabi-Yaus, or orientifolds of Calabi-Yau hypersurfaces in toric varieties \cite{Kreuzer:2000xy}. Select one such manifold, $X$, and compute the triple intersection numbers and the K\"ahler cone of $X$.  At a point in K\"ahler moduli space that lies well inside the K\"ahler cone, the kinetic matrix $K_{ij}$ for $C_4$ axions is determined by the leading-order K\"ahler potential, which involves only the triple intersection numbers and the K\"ahler moduli vevs, and so the axion kinetic couplings are fully specified.

\vskip 3pt
Next, to compute the axion potential, one needs to identify four-cycles in $X$ that have the correct number of fermion zero modes to support Euclidean D3-brane contributions to the superpotential \cite{Witten:1996bn} or the K\"ahler potential.  At present only a small subset of superpotential terms are well-understood.  Moreover, the axion charges of possible superpotential terms are constrained to lie inside a subregion of the charge space, corresponding to the cone of effective divisors in $H_4(X,\mathbb{Z})$, and so even a complete understanding of the superpotential would not suffice to test criteria such as the LWGC that involve the full charge lattice.
Charges outside the cone of effective divisors could be carried by non-BPS Euclidean D3-branes wrapping non-calibrated four-cycles, and these could contribute to the K\"ahler potential, or to higher F-terms \cite{Beasley:2005iu}.  However, hardly anything is known about such non-BPS instantons --- see for example \cite{Demirtas:2019lfi}.

\vskip 3pt
Even if one could compute or model all the significant instanton contributions to the effective action, a remaining limitation is that the properties of the axion potential vary from one compactification to another, and within a single compactification they vary from point to point in the moduli space.  There is no canonical measure for sampling the resulting array of effective theories.  For this reason it is more feasible to investigate which axion theories \emph{can arise} in string compactifications than it is to discuss the relative rarity of different classes of theories.  Even for the broadest properties, such as the number of axions, few robust statements about relative likelihood are now available.

\subsubsection{Hierarchical axion couplings}

One general finding \cite{Demirtas:2018akl} is that within the region of control of the $\alpha'$ expansion, there are large hierarchies in the sizes of cycles in $X$.
This phenomenon can be traced to the fact that the K\"ahler cone of $X$ typically has a small opening angle when $h^{1,1}(X) \gg 1$.
The hierarchies in cycle sizes lead to hierarchies in the kinetic term and, more dramatically, in the magnitudes of nonperturbative  superpotential terms corresponding to Euclidean D3-branes wrapping different cycles.

\vskip 3pt
One might hope that for sufficiently large $h^{1,1}$, matrices such as the kinetic matrix $K_{ij}$ can be well-approximated as elements of suitable random matrix ensembles, such as the Wishart and inverse-Wishart models examined in \S\ref{sec:rmt}.
The utility of such modeling depends on how large $h^{1,1}$ needs to be for a random matrix description to be accurate, and on whether the random matrix ensemble is simple enough to be well-understood.  A rather optimistic guess would be that Gaussian (i.e., Wigner) and Wishart ensembles, and close relatives, could give useful models starting around $h^{1,1} \sim 10-20$.
The reality is more challenging: correlations in the underlying geometric structures, and heavy-tailed eigenvalue distributions \cite{Long:2014fba} stemming from the hierarchy in cycle sizes noted above, have precluded precise matches to simple random matrix models, at least in the regime $h^{1,1}(X) \lesssim 30$ that has been well-studied.  It remains possible that universality will take hold when the number of axions is much larger, of order hundreds.

\subsubsection{Alignment}

As noted above, a complete accounting of all instanton terms in $\mathcal{N}=1$ compactifications is a distant goal, so any attempts to exhibit alignment or clockwork models in string theory are necessarily preliminary.

\vskip 3pt
Alignment in compactifications on Calabi-Yau hypersurfaces with $h^{1,1} \le 4$ was considered in
\cite{Long:2016jvd}, via a simplified model of the charge matrix $Q_{ij}$.  No significant alignment was found.
For larger $h^{1,1}$ the charge matrix can in principle support a large degree of alignment, but at the same time the typical compactification volume necessary for control of the $\alpha'$ expansion grows as a power of $h^{1,1}$.  The resulting reduction of the scale of the kinetic term, and so of the axion periodicities in Planck units, often erases any gains due to alignment.

\vskip 3pt
While it is possible that super-Planckian field ranges due to KNP alignment can occur in Calabi-Yau hypersurfaces with $h^{1,1} \gg 1$, this appears unlikely to be a common feature.

\subsubsection{Clockwork and sequestering}\label{sec:clockworkapp}

Clockwork structures seem contrived from the viewpoint of the axion charge matrices seen in Calabi-Yau compactifications of string theory.  However, one of the original arguments in favor of clockwork is largely bottom-up, relying on locality in an extra dimension \cite{Kaplan:2015fuy}, and one might be tempted to expect an ultraviolet completion in string theory, for example involving an array of D-branes spaced out around an extra dimension.

\vskip 3pt
We would like to point out that, even setting aside the particular symmetry structures required in \cite{Kaplan:2015fuy}, it is not clear that the requisite degree of sequestering of sectors is possible in a string compactification.  The difficulty is that unwanted couplings between spatially-separated sectors can be induced by light fields propagating in the bulk of the extra dimension(s).  If all charged fields in the bulk had masses at least as large as the Kaluza-Klein scale, then Yukawa suppression of their propagators would ensure adequate sequestering, as proposed in \cite{Randall:1998uk}.  This sort of extra-dimensional locality is a critical assumption made in \cite{Kaplan:2015fuy}, and the question now is whether it arises in string theory.

\vskip 3pt
In the context of the mediation of supersymmetry breaking, sequestering of one sector from another via separation along an extra dimension has \emph{not} been convincingly established in string theory.  The essential difficulty is that string compactifications involve light moduli, as well as light states of stretched open strings, that are not heavy enough to be neglected.  In particular, scenarios based on warped sequestering \cite{Kachru:2007xp}, which is the gravity dual of conformal sequestering \cite{Schmaltz:2006qs}, have been shown \cite{Berg:2010ha,Berg:2012aq} to be spoiled by moduli-mediated effects.  This finding is not surprising from the viewpoint of inflationary model-building: it is an incarnation of the $\eta$ problem.\footnote{See Chapter 4 of \cite{Baumann:2014nda} for a summary.}

\vskip 3pt
We conclude that the proposed mechanisms for realizing clockwork in string theory face significant obstacles.  While we have no grounds for suggesting that clockwork cannot arise in string theory, we do expect such models to be rare at best.

\section{Saddle Point Approximation for Isotropic N-flation}\label{sec:saddle}

In this appendix, we supplement the analysis of \S\ref{ssec:isotropic} by computing the variance of the $n$-th harmonic in an isotropic $N$-flation model by means of a saddle point approximation. The variance takes the form
\begin{equation}
  \sigma_n^2 = \sum_s k^{(N, n)}_s e^{- 2 \alpha \sqrt{s}}\,,
\end{equation}
where $k_s^{(N, n)}$ is a multiplicity factor that counts the number of
lattice sites satisfying:
\begin{equation}
  \sum_i \ell_i = n \;, \qquad \sum_i \ell_i^2 = s\;.
\end{equation}
Note that
\begin{equation}
  s \pm n = \sum_i \ell_i  (\ell_i \pm 1) \in 2\mathbbm{Z}_{\geqslant 0}\;,
\end{equation}
so $s - | n |$ is non-negative and even.

\vskip 3pt
The multiplicity factor that appears is a sum of multinomial coefficients:
\begin{equation}
  k^{(N, n)}_s = \sum_{\{ r_a \}} \frac{N!}{\prod_a r_a !}\;, \label{eqn:ksum}
\end{equation}
where $r_a \equiv | \{ i|\ell_i = a \} |$ denotes the number of $a$'s
appearing in $\vec{\ell}$, and the sum is restricted by the conditions
\begin{equation}
  r_a \geqslant 0 \;, \qquad \sum_a r_a = N \;, \qquad \sum_a a r_a = n \;,
  \qquad \sum_a a^2 r_a = s\;.
\end{equation}
We estimate $k^{(N, n)}_s$ by finding the saddle point of this
multidimensional sum. Let $\hat{r}_a$ denote the location of the saddle. We
then have:
\begin{equation}
  \frac{N!}{\prod_a r_a !} = e^{- S} \;, \qquad S = \hat{S} + \sum_a (r_a -
  \hat{r}_a) \psi (\hat{r}_a + 1) + \frac{1}{2}  \sum_a (r_a - \hat{r}_a)^2
  \psi^{(1)} (\hat{r}_a + 1) + \ldots
\end{equation}
Here the deviations $\delta_a \equiv r_a - \hat{r}_a$ are constrained by
$\sum_a a^p \delta_a = 0$ for $p = 0, 1, 2$. The saddle-point conditions
$\sum_a \delta_a \psi (\hat{r}_a + 1) = 0$ have the solution:
\begin{equation}
  \psi (\hat{r}_a + 1) = \log M + a \log w + a^2 \log q\;,
\end{equation}
where the constants $M$, $w$ and $q$ will determine $N, n$ and $s$.

\vskip 3pt
Having found the saddle point, we now consider the integral near the saddle
point. We have
\begin{equation}
  S = \hat{S} + \frac{1}{2}  \sum_a \delta_a^2 \psi^{(1)} (\hat{r}_a + 1) +
  \ldots
\end{equation}
where the $\delta_a$ are constrained by $\sum_a a^p \delta_a = 0$ for $p = 0,
1, 2$, as above. To evaluate the determinant in the Gaussian integral, we
start be considering the simpler problem of evaluating the determinant for
\begin{equation}
  S = \hat{S} + \frac{1}{2}  \sum_a \Delta_a^2 \psi^{(1)} (\hat{r}_a + 1) +
  \ldots
\end{equation}
where the $\Delta_a$ are unconstrained. From this, we obtain the rough
estimate:
\begin{equation}
  \sum_{\{ r_a \}} \frac{N!}{\prod_a r_a !} \sim N! \prod_a \sqrt{\frac{2
  \pi}{\psi^{(1)} (\hat{r}_a + 1)}} \cdot \frac{1}{\Gamma (\hat{r}_a + 1)}\;.
  \label{eqn:naiveest}
\end{equation}
However, this is incorrect because we have integrated over three extra
variables. To account for this, we make the change of variables $\Delta_a
\rightarrow u_b$,
\begin{equation}
  \Delta_a = \sum_{| b | \leqslant 1} u_b A^b_{\; a} + \sum_{| b | > 1} u_b
  B^b_{\; a}\;,
\end{equation}
where
\begin{align}
  A^b_{\; a} &= \frac{1}{\psi^{(1)} (\hat{r}_a + 1)} \sum_{p = 0}^2
  \lambda^b_{\; p} a^p \;, \nonumber \\ B^b_{\; - 1, 0, 1} &= \left\{ - \frac{b (b -
  1)}{2}, b^2 - 1, - \frac{b (b + 1)}{2} \right\} \;, \\ B^b_b &= 1\;.
\end{align}
Here $\lambda^b_{\; p}$ is chosen such that
\begin{equation}
  \sum_a A^b_{\; a} a^p = b^p \;, \qquad (p = 0, 1, 2) \;, \label{eqn:Acons}
\end{equation}
This ensures that the Jacobian for the change of coordinates is one. In
particular, changing variables from $u_a$ to $\hat{u}_a = \Delta_a$ for $| a |
> 1$ (keeping $u_a = \hat{u}_a$ for $| a | \leqslant 1$) gives a triangular
Jacobian matrix with ones on the diagonal. We have
\begin{equation}
  \Delta_a = \sum_{| b | \leqslant 1} u_b  \hat{A}^b_{\; a} + \sum_{| b | > 1}
  \hat{u}_b B^b_{\; a}\;,
\end{equation}
where $\hat{A}^b_{\; a}$ is now a $3 \times 3$ matrix, which is fixed by the
observation that
\begin{equation}
  \sum_a \hat{A}^b_{\; a} a^p = \sum_a A^b_{\; a} a^p = b^p \;, \qquad (p = 0,
  1, 2) \;,
\end{equation}
since $\sum_a B^b_{\; a} a^p = 0$ for $p = 0, 1, 2$ by construction. This
implies that $\hat{A}^b_{\; a} = \delta^b_{\; a}$, hence the Jacobian for the
change between $\Delta_a$ and $\hat{u}_a$ is again triangular with ones on the
diagonal.

\vskip 3pt
The constraints $\sum_a a^p \delta_a = 0$ for $p = 0, 1, 2$ on the physical
integration variables are equivalent to $u_a = 0$ for $a = 0, \pm 1$.
Moreover, we have
\begin{equation}
  S = \hat{S} + \frac{1}{2}  \sum_a \psi^{(1)} (\hat{r}_a + 1) \sum_{| b |, |
  b' | \leqslant 1} A_a^b A^{b'}_{\; a} u_b u_{b'} + \frac{1}{2}  \sum_a
  \psi^{(1)} (\hat{r}_a + 1) \sum_{| b |, | b' | > 1} B_a^b B^{b'}_{\; a} u_b
  u_{b'}\;,
\end{equation}
since the cross terms vanish by construction. Thus, the correct result can be
obtained from the estimate (\ref{eqn:naiveest}) by dividing by the Gaussian
integral over the extra variables $u_{0, \pm 1}$:
\begin{equation}
  \sum_{\{ r_a \}} \frac{N!}{\prod_a r_a !} \simeq N! \sqrt{\frac{\det
  \mathbbm{A}^{b b'}}{(2 \pi)^3}}  \prod_a \sqrt{\frac{2 \pi}{\psi^{(1)}
  (\hat{r}_a + 1)}} \cdot \frac{1}{\Gamma (\hat{r}_a + 1)}\;,
  \label{eqn:correctsaddle}
\end{equation}
where $\mathbbm{A}^{b b'}$ is the $3 \times 3$ matrix
\begin{equation}
  \mathbbm{A}^{b b'} = \sum_a A_a^b A^{b'}_{\; a} \psi^{(1)} (\hat{r}_a + 1)\;.
\end{equation}
To evaluate this determinant, note that the constraint (\ref{eqn:Acons}) can
be rewritten as:
\begin{equation}
  \sum_{p = 0}^2 \lambda^b_{\; p} \beta_{p + r} = b^r \;, \qquad (r = 0, 1, 2)
  \;,
\end{equation}
where
\begin{equation}
  \beta_p \equiv \sum_a \frac{a^p}{\psi^{(1)} (\hat{r}_a + 1)}\;.
  \label{eqn:betamat}
\end{equation}
In this notation, we have as well
\begin{equation}
  \mathbbm{A}^{b b'} = \sum_{p, p' = 0}^2 \lambda^b_{\; p} \lambda^{b'}_{\; p'}
  \beta_{p + p'}\;.
\end{equation}
Thus,
\begin{equation}
  \det \mathbbm{A}= \left( \det \lambda^b_{\; p} \right)^2  (\det \beta_{p +
  p'}) = \frac{4}{\det \beta_{p + p'}}\;,
\end{equation}
where indices $b$ run over $- 1, 0, 1$, $p, p'$ run over $0, 1, 2$, and we use
$\det b^p = 2$.

\vskip 3pt
To further simplify the above result, we use Stirling's approximation for
$\hat{r}_a \gg 1$. We have
\begin{equation}
  \psi (\hat{r}_a + 1) \simeq \log \left[ \hat{r}_a + \frac{1}{2} + O \left(
  \frac{1}{\hat{r}_a} \right) \right]\;,
\end{equation}
so that
\begin{equation}
  \hat{r}_a \simeq M q^{a^2} w^a - \frac{1}{2}\;.
\end{equation}
This implies in particular that
\begin{equation}
  N \simeq \sum'_a \left[ M q^{a^2} w^a - \frac{1}{2} \right] \;, \qquad n
  \simeq \sum'_a a \left[ M q^{a^2} w^a - \frac{1}{2} \right] \;, \qquad s
  \simeq \sum'_a a^2  \left[ M q^{a^2} w^a - \frac{1}{2} \right]\;.
  \label{eqn:Nns}
\end{equation}
Here we indicate with a prime that these sums must be cut off. The same cutoff
should be used for (\ref{eqn:correctsaddle}) and (\ref{eqn:betamat}) above. A
reasonable cutoff is $\hat{r}_a \gtrsim 0$ (where we assume that $| q | < 1$ to
make the result sensible); as long as the same cutoff is used for each
sum/product, the result should not greatly depend on the exact choice of
cutoff.

\vskip 3pt
We likewise have:
\begin{align}
  \psi^{(1)} (\hat{r}_a + 1) &\simeq \frac{1}{\hat{r}_a + \frac{1}{2}} \simeq
  \frac{1}{M q^{a^2} w^a}\;, \nonumber \\
  \Gamma (\hat{r}_a + 1) &\simeq \sqrt{2 \pi}  \left( \hat{r}_a + \frac{1}{2}
  \right)^{\hat{r}_a + \frac{1}{2}} e^{- \left( \hat{r}_a + \frac{1}{2}
  \right)} \simeq \sqrt{2 \pi}  (M q^{a^2} w^a)^{M q^{a^2} w^a} e^{- M q^{a^2}
  w^a} \;.
\end{align}
Working everything out, we find:
\begin{equation}
\boxed{k^{(N, n)}_s \simeq \frac{N^{N + \frac{1}{2}} e^{\frac{a_{\max} - a_{\min} +
   1}{2}}}{\pi M^N q^s w^n  \sqrt{\det_{0 \leqslant p, p' \leqslant 2}
   \beta_{p + p'}}}} \label{eqn:ksSaddle0}
\end{equation}
where we apply $N! \simeq \sqrt{2 \pi N} N^N e^{- N}$.

\vskip 3pt
The explicit dependence on $M$ and $a_{\max} - a_{\min}$ can be eliminated as
follows. Define
\begin{equation}
  \theta_p \equiv \beta_p / M = \sum'_a a^p q^{a^2} w^a\;.
\end{equation}
We then have,
\begin{equation}
  N \simeq M \theta_0 - \frac{a_{\max} - a_{\min} + 1}{2} \qquad
  \Longrightarrow \qquad \frac{N}{M} \simeq \frac{\theta_0}{1 + \frac{a_{\max}
  - a_{\min} + 1}{2 N}}\;.
\end{equation}
Provided that $a_{\max} - a_{\min} + 1 \ll \sqrt{N}$, (\ref{eqn:ksSaddle0})
simplifies to
\begin{equation}
  k^{(N, n)}_s \simeq \frac{\theta_0^N}{\pi N q^s w^n}  \left[ \det_{0
  \leqslant p, p' \leqslant 2}  \frac{\theta_{p + p'}}{\theta_0} \right]^{- 1
  / 2}. \label{eqn:ksSaddle}
\end{equation}

\vskip 3pt
The above saddle point calculation can be simplified if we drop all but the
leading terms in the $N \rightarrow \infty$ limit with $q$ and $w$ fixed. This amounts to removing the cutoff and the $-1/2$ constant term from the sums in (\ref{eqn:Nns}) and ignoring the term in brackets in (\ref{eqn:ksSaddle}). Thus we find
\begin{align}
  \frac{1}{N} \log k_s^{(N, n)} &\simeq \log \theta (w ; q) - \sigma \log q -
  \nu \log w + O \left( \frac{\log N}{N} \right)\;, \nonumber \\
  \sigma \equiv \frac{s}{N} &\simeq \frac{\theta^{(2)} (w ; q)}{\theta (w ;
  q)} + O \left( \frac{(\log N)^{3 / 2}}{N} \right)\;, \nonumber \\
  \nu \equiv \frac{n}{N} &\simeq \frac{\theta^{(1)} (w ; q)}{\theta (w ; q)}
  + O \left( \frac{\log N}{N} \right) \;, \label{eq:ksLeadingLargeN}
\end{align}
where we used $a_{\text{cut}} = O \left( \sqrt{\log N} \right)$, as above, and
\begin{equation}
  \theta^{(p)} (w ; q) \equiv \sum_{a = - \infty}^{\infty} a^p q^{a^2} w^a\,,
\end{equation}
with $\theta (w ; q)$ the Jacobi theta function and $\theta^{(p)} (w ; q)$ its
derivatives.

\vskip 3pt
Note that (\ref{eq:ksLeadingLargeN}) takes the form of a thermodynamic
system. In particular, note that
\begin{equation}
  \sigma = \frac{\partial}{\partial \log q} \log \theta (w ; q)\;, \qquad \qquad
  \nu = \frac{\partial}{\partial \log w} \log \theta (w ; q)\;.
\end{equation}
Thus, if we treat $N \log \theta (w ; q)$ as a thermodynamic potential with
the natural variables $\log w$ and $\log q$, then $\log k_s^{(N, n)}$ is the
related potential with conjugate natural variables $s = N \sigma$ and $n = N
\nu$, related by the usual Legendre transform:
\begin{equation}
  \log k_s^{(N, n)} = N \log \theta (w ; q) - s \log q - n \log w\;.
\end{equation}
In this way, we can think of $\log q$ and $\log w$ as chemical potentials
associated to $s$ and $n$, with associated thermodynamic potential $N \log
\theta (w ; q)$.\footnote{We can also treat $N$ and $\log M$ as
thermodynamically conjugate variables, as follows. Applying Stirling's approximation, we have $\log \frac{k_s^{(N, n)}}{N!} = N - N \log M - s \log q - n \log w$. The potential $F = N = M \theta (w ; q)$ satisfies $\frac{\partial
F}{\partial \log M} = N$, $\frac{\partial F}{\partial \log w} = n$, and
$\frac{\partial F}{\partial \log q} = s$, so $\log \frac{k_s^{(N, n)}}{N!}$ is
the Legendre transform, similar to before. Moreover, $F - N \log M = N \log
\theta - \log N!$, so switching variables from $M$ to $N$ gives us back the
thermodynamic potential $N \log \theta$ considered above, up to a $q$ and $w$
independent additive factor.}

\vskip 3pt
We consider two special limits. In the limit $q w^{\pm 1} \ll
1$, we can approximate:
\begin{equation}
  \sigma \simeq \frac{q (w + w^{- 1})}{1 + q (w + w^{- 1})} \;, \qquad \nu
  \simeq \frac{q (w - w^{- 1})}{1 + q (w + w^{- 1})} \;,
\end{equation}
so that
\begin{equation}
  w \simeq \sqrt{\frac{\sigma + \nu}{\sigma - \nu}} \;, \qquad q \simeq
  \frac{1}{1 - \sigma}  \sqrt{\frac{\sigma^2 - \nu^2}{4}} \;.
\end{equation}
Using this, we obtain:
\begin{equation}
  \frac{1}{N} \log k_s^{(N, n)} \simeq - \frac{\sigma + \nu}{2} \log
  \frac{\sigma + \nu}{2} - \frac{\sigma - \nu}{2} \log \frac{\sigma - \nu}{2}
  - (1 - \sigma) \log (1 - \sigma) + O (\sigma^4 \log \sigma)\;.
  \label{eqn:Qpm1LargeN}
\end{equation}
This is exactly what one obtains from the approximation in
(\ref{eqn:ksLeadingSmallS}):
\begin{equation}
  k_s^{(N, n)} \sim \left(\begin{array}{c}
    N\\
    \frac{s + n}{2}, \frac{s - n}{2}, N-s
  \end{array}\right) \,, \label{eqn:Qpm1Approx}
\end{equation}
which comes from counting instantons with $\ell_a \in \{ 0, \pm 1 \}$.

\vskip 3pt
Let us instead consider the limit $q w^{\pm 1} \rightarrow 1$. In this case,
it is appropriate to apply the S-duality transformation:
\begin{equation}
  \vartheta (\zeta ; \tau) = \frac{e^{- \frac{\pi i \zeta^2}{\tau}}}{\sqrt{- i
  \tau}} \vartheta \left( \frac{\zeta}{\tau}, - \frac{1}{\tau} \right)\;,
\end{equation}
where we write $\vartheta (\zeta ; \tau) = \theta (e^{2 \pi i \zeta} ; e^{\pi
i \tau})$. Let us redefine $t = - i \tau$ and $z = - i \zeta$, so that
\begin{equation}
  \vartheta (i z ; i t) = \frac{e^{\frac{\pi z^2}{t}}}{\sqrt{t}} \vartheta
  \left( \frac{z}{t}, \frac{i}{t} \right)\;.
\end{equation}
We have $\vartheta^{(p)} (\zeta ; \tau) \equiv \theta^{(p)} (e^{2 \pi i \zeta}
; e^{\pi i \tau}) = \frac{1}{(2 \pi i)^p} \partial_{\zeta}^p \vartheta (\zeta
; \tau)$, so that
\begin{align}
  \vartheta^{(1)} (i z, i t) &= - \frac{e^{\frac{\pi z^2}{t}}}{t^{3 / 2}}
  \left[ z \vartheta \left( \frac{z}{t}, \frac{i}{t} \right) + i
  \vartheta^{(1)} \left( \frac{z}{t}, \frac{i}{t} \right) \right]\;, \nonumber \\
  \vartheta^{(2)} (i z, i t) &=  \frac{e^{\frac{\pi z^2}{t}}}{t^{5 / 2}}
  \left[ z^2 \vartheta \left( \frac{z}{t}, \frac{i}{t} \right) + 2 i z
  \vartheta^{(1)} \left( \frac{z}{t}, \frac{i}{t} \right) + \frac{t}{2 \pi}
  \vartheta \left( \frac{z}{t}, \frac{i}{t} \right) - \vartheta^{(2)} \left(
  \frac{z}{t}, \frac{i}{t} \right) \right] \;.
\end{align}

\vskip 3pt
From this, we obtain:
\begin{equation}
  \sigma \simeq \frac{1}{t^2}  \left[ z^2 + \frac{t}{2 \pi} + 2 i z
  \frac{\vartheta^{(1)} \left( \frac{z}{t}, \frac{i}{t} \right)}{\vartheta
  \left( \frac{z}{t}, \frac{i}{t} \right)} - \frac{\vartheta^{(2)} \left(
  \frac{z}{t}, \frac{i}{t} \right)}{\vartheta \left( \frac{z}{t}, \frac{i}{t}
  \right)} \right] \;, \qquad \nu \simeq - \frac{1}{t}  \left[ z + i
  \frac{\vartheta^{(1)} \left( \frac{z}{t}, \frac{i}{t} \right)}{\vartheta
  \left( \frac{z}{t}, \frac{i}{t} \right)} \right]\;.
\end{equation}
We have
\begin{equation}
  \vartheta^{(p)} \left( \frac{z}{t}, \frac{i}{t} \right) \simeq \delta_{p, 0}
  + 2 e^{- \frac{\pi}{t}}  \left\{ \begin{array}{ccc}
    \cos \left( \frac{2 \pi z}{t} \right) &  & \text{$p$ even}\\
    - \sin \left( \frac{2 \pi z}{t} \right) &  & \text{$p$ odd}
  \end{array} \right.\,.
\end{equation}
Because of the leading $e^{- \frac{\pi}{t}}$, in the $t \rightarrow 0$ limit
we can approximate:
\begin{equation}
  \sigma \simeq \left( \frac{z}{t} \right)^2 + \frac{1}{2 \pi t} \;, \qquad
  \nu \simeq - \frac{z}{t}\;.
\end{equation}
Hence
\begin{equation}
  t \simeq \frac{1}{2 \pi (\sigma - \nu^2)} \;, \qquad z \simeq - \frac{\nu}{2
  \pi (\sigma - \nu^2)}\;.
\end{equation}
We then find:
\begin{align}
  \frac{1}{N} \log k_s^{(N, n)} \simeq \frac{1}{2} \log t^{- 1} + \frac{\pi
  z^2}{t} + \pi \sigma t + 2 \pi \nu z & \simeq \frac{1}{2} \log [2 \pi e
  (\sigma - \nu^2)] \nonumber \\
 & \simeq \frac{1}{2} \log \left[\frac{2 \pi e}{N} \left(s - \frac{n^2}{N} \right)\right]\;,
  \label{eqn:contLargeN}
\end{align}
which matches the continuum limit (\ref{eqn:continuumLimit}).

\newpage

\bibliographystyle{JHEP}
 \bibliography{ref}

\providecommand{\href}[2]{#2}\begingroup\raggedright\begin{thebibliography}{10}

\bibitem{Akrami:2018odb}
{\scshape Planck} collaboration, Y.~Akrami et~al., \emph{{Planck 2018 results.
  X. Constraints on inflation}},  \href{http://arxiv.org/abs/1807.06211}{{\tt
  1807.06211}}.

\bibitem{Abazajian:2016yjj}
{\scshape CMB-S4} collaboration, K.~N. Abazajian et~al., \emph{{CMB-S4 Science
  Book, First Edition}},  \href{http://arxiv.org/abs/1610.02743}{{\tt
  1610.02743}}.

\bibitem{banks:2003sx}
T.~Banks, M.~Dine, P.~J. Fox and E.~Gorbatov, \emph{{On the possibility of
  large axion decay constants}},
  \href{http://dx.doi.org/10.1088/1475-7516/2003/06/001}{\emph{JCAP} {\bf 0306}
  (2003) 001}, [\href{http://arxiv.org/abs/hep-th/0303252}{{\tt
  hep-th/0303252}}].

\bibitem{Arkanihamed:2006dz}
N.~Arkani-Hamed, L.~Motl, A.~Nicolis and C.~Vafa, \emph{{The String landscape,
  black holes and gravity as the weakest force}},
  \href{http://dx.doi.org/10.1088/1126-6708/2007/06/060}{\emph{JHEP} {\bf 0706}
  (2007) 060}, [\href{http://arxiv.org/abs/hep-th/0601001}{{\tt
  hep-th/0601001}}].

\bibitem{silverstein:2008sg}
E.~Silverstein and A.~Westphal, \emph{{Monodromy in the CMB: Gravity Waves and
  String Inflation}},
  \href{http://dx.doi.org/10.1103/PhysRevD.78.106003}{\emph{Phys.Rev.} {\bf
  D78} (2008) 106003}, [\href{http://arxiv.org/abs/0803.3085}{{\tt
  0803.3085}}].

\bibitem{mcallister:2008hb}
L.~McAllister, E.~Silverstein and A.~Westphal, \emph{{Gravity Waves and Linear
  Inflation from Axion Monodromy}},
  \href{http://dx.doi.org/10.1103/PhysRevD.82.046003}{\emph{Phys.Rev.} {\bf
  D82} (2010) 046003}, [\href{http://arxiv.org/abs/0808.0706}{{\tt
  0808.0706}}].

\bibitem{liddle:1998jc}
A.~R. Liddle, A.~Mazumdar and F.~E. Schunck, \emph{{Assisted inflation}},
  \href{http://dx.doi.org/10.1103/PhysRevD.58.061301}{\emph{Phys. Rev.} {\bf
  D58} (1998) 061301}, [\href{http://arxiv.org/abs/astro-ph/9804177}{{\tt
  astro-ph/9804177}}].

\bibitem{cheung:2014vva}
C.~Cheung and G.~N. Remmen, \emph{{Naturalness and the Weak Gravity
  Conjecture}},
  \href{http://dx.doi.org/10.1103/PhysRevLett.113.051601}{\emph{Phys.Rev.Lett.}
  {\bf 113} (2014) 051601}, [\href{http://arxiv.org/abs/1402.2287}{{\tt
  1402.2287}}].

\bibitem{Heidenreich:2016aqi}
B.~Heidenreich, M.~Reece and T.~Rudelius, \emph{{Evidence for a Lattice Weak
  Gravity Conjecture}},
  \href{http://dx.doi.org/10.1007/JHEP08(2017)025}{\emph{JHEP} {\bf 08} (2017)
  025}, [\href{http://arxiv.org/abs/1606.08437}{{\tt 1606.08437}}].

\bibitem{Montero:2016tif}
M.~Montero, G.~Shiu and P.~Soler, \emph{{The Weak Gravity Conjecture in three
  dimensions}}, \href{http://dx.doi.org/10.1007/JHEP10(2016)159}{\emph{JHEP}
  {\bf 10} (2016) 159}, [\href{http://arxiv.org/abs/1606.08438}{{\tt
  1606.08438}}].

\bibitem{Valenzuela:2016yny}
I.~Valenzuela, \emph{{Backreaction Issues in Axion Monodromy and Minkowski
  4-forms}}, \href{http://dx.doi.org/10.1007/JHEP06(2017)098}{\emph{JHEP} {\bf
  06} (2017) 098}, [\href{http://arxiv.org/abs/1611.00394}{{\tt 1611.00394}}].

\bibitem{Heidenreich:2017sim}
B.~Heidenreich, M.~Reece and T.~Rudelius, \emph{{The Weak Gravity Conjecture
  and Emergence from an Ultraviolet Cutoff}},
  \href{http://arxiv.org/abs/1712.01868}{{\tt 1712.01868}}.

\bibitem{Lee:2018urn}
S.-J. Lee, W.~Lerche and T.~Weigand, \emph{{Tensionless Strings and the Weak
  Gravity Conjecture}},
  \href{http://dx.doi.org/10.1007/JHEP10(2018)164}{\emph{JHEP} {\bf 10} (2018)
  164}, [\href{http://arxiv.org/abs/1808.05958}{{\tt 1808.05958}}].

\bibitem{Lee:2018spm}
S.-J. Lee, W.~Lerche and T.~Weigand, \emph{{A Stringy Test of the Scalar Weak
  Gravity Conjecture}},
  \href{http://dx.doi.org/10.1016/j.nuclphysb.2018.11.001}{\emph{Nucl. Phys.}
  {\bf B938} (2019) 321--350}, [\href{http://arxiv.org/abs/1810.05169}{{\tt
  1810.05169}}].

\bibitem{Lee:2019tst}
S.-J. Lee, W.~Lerche and T.~Weigand, \emph{{Modular Fluxes, Elliptic Genera,
  and Weak Gravity Conjectures in Four Dimensions}},
  \href{http://dx.doi.org/10.1007/JHEP08(2019)104}{\emph{JHEP} {\bf 08} (2019)
  104}, [\href{http://arxiv.org/abs/1901.08065}{{\tt 1901.08065}}].

\bibitem{Andriolo:2018lvp}
S.~Andriolo, D.~Junghans, T.~Noumi and G.~Shiu, \emph{{A Tower Weak Gravity
  Conjecture from Infrared Consistency}},
  \href{http://dx.doi.org/10.1002/prop.201800020}{\emph{Fortsch. Phys.} {\bf
  66} (2018) 1800020}, [\href{http://arxiv.org/abs/1802.04287}{{\tt
  1802.04287}}].

\bibitem{Heidenreich:2015nta}
B.~Heidenreich, M.~Reece and T.~Rudelius, \emph{{Sharpening the Weak Gravity
  Conjecture with Dimensional Reduction}},
  \href{http://dx.doi.org/10.1007/JHEP02(2016)140}{\emph{JHEP} {\bf 02} (2016)
  140}, [\href{http://arxiv.org/abs/1509.06374}{{\tt 1509.06374}}].

\bibitem{dimopoulos:2005ac}
S.~Dimopoulos, S.~Kachru, J.~McGreevy and J.~G. Wacker, \emph{{N-flation}},
  \href{http://dx.doi.org/10.1088/1475-7516/2008/08/003}{\emph{JCAP} {\bf 0808}
  (2008) 003}, [\href{http://arxiv.org/abs/hep-th/0507205}{{\tt
  hep-th/0507205}}].

\bibitem{kim:2004rp}
J.~E. Kim, H.~P. Nilles and M.~Peloso, \emph{{Completing natural inflation}},
  \href{http://dx.doi.org/10.1088/1475-7516/2005/01/005}{\emph{JCAP} {\bf 0501}
  (2005) 005}, [\href{http://arxiv.org/abs/hep-ph/0409138}{{\tt
  hep-ph/0409138}}].

\bibitem{Choi:2014rja}
K.~Choi, H.~Kim and S.~Yun, \emph{{Natural inflation with multiple
  sub-Planckian axions}},
  \href{http://dx.doi.org/10.1103/PhysRevD.90.023545}{\emph{Phys. Rev.} {\bf
  D90} (2014) 023545}, [\href{http://arxiv.org/abs/1404.6209}{{\tt
  1404.6209}}].

\bibitem{Choi:2015fiu}
K.~Choi and S.~H. Im, \emph{{Realizing the relaxion from multiple axions and
  its UV completion with high scale supersymmetry}},
  \href{http://dx.doi.org/10.1007/JHEP01(2016)149}{\emph{JHEP} {\bf 01} (2016)
  149}, [\href{http://arxiv.org/abs/1511.00132}{{\tt 1511.00132}}].

\bibitem{Kaplan:2015fuy}
D.~E. Kaplan and R.~Rattazzi, \emph{{Large field excursions and approximate
  discrete symmetries from a clockwork axion}},
  \href{http://dx.doi.org/10.1103/PhysRevD.93.085007}{\emph{Phys. Rev.} {\bf
  D93} (2016) 085007}, [\href{http://arxiv.org/abs/1511.01827}{{\tt
  1511.01827}}].

\bibitem{freese:1990rb}
K.~Freese, J.~A. Frieman and A.~V. Olinto, \emph{{Natural inflation with pseudo
  - Nambu-Goldstone bosons}},
  \href{http://dx.doi.org/10.1103/PhysRevLett.65.3233}{\emph{Phys.Rev.Lett.}
  {\bf 65} (1990) 3233--3236}.

\bibitem{Pimentel:2019otp}
G.~L. Pimentel and J.~Stout, \emph{{Real-Time Corrections to the Effective
  Potential}},  \href{http://arxiv.org/abs/1905.00219}{{\tt 1905.00219}}.

\bibitem{Candelas:1990rm}
P.~Candelas, X.~C. De~La~Ossa, P.~S. Green and L.~Parkes, \emph{{A Pair of
  Calabi-Yau manifolds as an exactly soluble superconformal theory}},
  \href{http://dx.doi.org/10.1016/0550-3213(91)90292-6}{\emph{Nucl. Phys.} {\bf
  B359} (1991) 21--74}.

\bibitem{Seiberg:1994rs}
N.~Seiberg and E.~Witten, \emph{{Electric - magnetic duality, monopole
  condensation, and confinement in N=2 supersymmetric Yang-Mills theory}},
  \href{http://dx.doi.org/10.1016/0550-3213(94)90124-4,
  10.1016/0550-3213(94)00449-8}{\emph{Nucl. Phys.} {\bf B426} (1994) 19--52},
  [\href{http://arxiv.org/abs/hep-th/9407087}{{\tt hep-th/9407087}}].

\bibitem{Nekrasov:2002qd}
N.~A. Nekrasov, \emph{{Seiberg-Witten prepotential from instanton counting}},
  \href{http://dx.doi.org/10.4310/ATMP.2003.v7.n5.a4}{\emph{Adv. Theor. Math.
  Phys.} {\bf 7} (2003) 831--864},
  [\href{http://arxiv.org/abs/hep-th/0206161}{{\tt hep-th/0206161}}].

\bibitem{Saraswat:2016eaz}
P.~Saraswat, \emph{{Weak gravity conjecture and effective field theory}},
  \href{http://dx.doi.org/10.1103/PhysRevD.95.025013}{\emph{Phys. Rev.} {\bf
  D95} (2017) 025013}, [\href{http://arxiv.org/abs/1608.06951}{{\tt
  1608.06951}}].

\bibitem{Furuuchi:2017upe}
K.~Furuuchi, \emph{{Weak Gravity Conjecture From Low Energy Observers'
  Perspective}},  \href{http://arxiv.org/abs/1712.01302}{{\tt 1712.01302}}.

\bibitem{Brown:2015iha}
J.~Brown, W.~Cottrell, G.~Shiu and P.~Soler, \emph{{Fencing in the Swampland:
  Quantum Gravity Constraints on Large Field Inflation}},
  \href{http://dx.doi.org/10.1007/JHEP10(2015)023}{\emph{JHEP} {\bf 10} (2015)
  023}, [\href{http://arxiv.org/abs/1503.04783}{{\tt 1503.04783}}].

\bibitem{Montero:2015ofa}
M.~Montero, A.~M. Uranga and I.~Valenzuela, \emph{{Transplanckian axions!?}},
  \href{http://dx.doi.org/10.1007/JHEP08(2015)032}{\emph{JHEP} {\bf 08} (2015)
  032}, [\href{http://arxiv.org/abs/1503.03886}{{\tt 1503.03886}}].

\bibitem{Hebecker:2016dsw}
A.~Hebecker, P.~Mangat, S.~Theisen and L.~T. Witkowski, \emph{{Can
  Gravitational Instantons Really Constrain Axion Inflation?}},
  \href{http://dx.doi.org/10.1007/JHEP02(2017)097}{\emph{JHEP} {\bf 02} (2017)
  097}, [\href{http://arxiv.org/abs/1607.06814}{{\tt 1607.06814}}].

\bibitem{Hebecker:2018ofv}
A.~Hebecker, T.~Mikhail and P.~Soler, \emph{{Euclidean wormholes, baby
  universes, and their impact on particle physics and cosmology}},
  \href{http://dx.doi.org/10.3389/fspas.2018.00035}{\emph{Front. Astron. Space
  Sci.} {\bf 5} (2018) 35}, [\href{http://arxiv.org/abs/1807.00824}{{\tt
  1807.00824}}].

\bibitem{Hertog:2018kbz}
T.~Hertog, B.~Truijen and T.~Van~Riet, \emph{{Euclidean axion wormholes have
  multiple negative modes}},
  \href{http://dx.doi.org/10.1103/PhysRevLett.123.081302}{\emph{Phys. Rev.
  Lett.} {\bf 123} (2019) 081302}, [\href{http://arxiv.org/abs/1811.12690}{{\tt
  1811.12690}}].

\bibitem{Hebecker:2019vyf}
A.~Hebecker and P.~Henkenjohann, \emph{{Gauge and gravitational instantons:
  From 3-forms and fermions to Weak Gravity and flat axion potentials}},
  \href{http://dx.doi.org/10.1007/JHEP09(2019)038}{\emph{JHEP} {\bf 09} (2019)
  038}, [\href{http://arxiv.org/abs/1906.07728}{{\tt 1906.07728}}].

\bibitem{Arkanihamed:2003wu}
N.~Arkani-Hamed, H.-C. Cheng, P.~Creminelli and L.~Randall, \emph{{Extra
  natural inflation}},
  \href{http://dx.doi.org/10.1103/PhysRevLett.90.221302}{\emph{Phys.Rev.Lett.}
  {\bf 90} (2003) 221302}, [\href{http://arxiv.org/abs/hep-th/0301218}{{\tt
  hep-th/0301218}}].

\bibitem{Delafuente:2014aca}
A.~de~la Fuente, P.~Saraswat and R.~Sundrum, \emph{{Natural Inflation and
  Quantum Gravity}},
  \href{http://dx.doi.org/10.1103/PhysRevLett.114.151303}{\emph{Phys.Rev.Lett.}
  {\bf 114} (2015) 151303}, [\href{http://arxiv.org/abs/1412.3457}{{\tt
  1412.3457}}].

\bibitem{Heidenreich:2015wga}
B.~Heidenreich, M.~Reece and T.~Rudelius, \emph{{Weak Gravity Strongly
  Constrains Large-Field Axion Inflation}},
  \href{http://dx.doi.org/10.1007/JHEP12(2015)108}{\emph{JHEP} {\bf 12} (2015)
  108}, [\href{http://arxiv.org/abs/1506.03447}{{\tt 1506.03447}}].

\bibitem{rudelius:2015xta}
T.~Rudelius, \emph{{Constraints on Axion Inflation from the Weak Gravity
  Conjecture}},
  \href{http://dx.doi.org/10.1088/1475-7516/2015/9/020}{\emph{JCAP} {\bf 09}
  (2015) 020}, [\href{http://arxiv.org/abs/1503.00795}{{\tt 1503.00795}}].

\bibitem{Brown:2015lia}
J.~Brown, W.~Cottrell, G.~Shiu and P.~Soler, \emph{{On Axionic Field Ranges,
  Loopholes and the Weak Gravity Conjecture}},
  \href{http://dx.doi.org/10.1007/JHEP04(2016)017}{\emph{JHEP} {\bf 04} (2016)
  017}, [\href{http://arxiv.org/abs/1504.00659}{{\tt 1504.00659}}].

\bibitem{junghans:2015hba}
D.~Junghans, \emph{{Large-Field Inflation with Multiple Axions and the Weak
  Gravity Conjecture}},
  \href{http://dx.doi.org/10.1007/JHEP02(2016)128}{\emph{JHEP} {\bf 02} (2016)
  128}, [\href{http://arxiv.org/abs/1504.03566}{{\tt 1504.03566}}].

\bibitem{Bachlechner:2015qja}
T.~C. Bachlechner, C.~Long and L.~McAllister, \emph{{Planckian Axions and the
  Weak Gravity Conjecture}},
  \href{http://dx.doi.org/10.1007/JHEP01(2016)091}{\emph{JHEP} {\bf 01} (2016)
  091}, [\href{http://arxiv.org/abs/1503.07853}{{\tt 1503.07853}}].

\bibitem{Hebecker:2015rya}
A.~Hebecker, P.~Mangat, F.~Rompineve and L.~T. Witkowski, \emph{{Winding out of
  the Swamp: Evading the Weak Gravity Conjecture with F-term Winding
  Inflation?}},
  \href{http://dx.doi.org/10.1016/j.physletb.2015.07.026}{\emph{Phys. Lett.}
  {\bf B748} (2015) 455--462}, [\href{http://arxiv.org/abs/1503.07912}{{\tt
  1503.07912}}].

\bibitem{Hebecker:2015zss}
A.~Hebecker, F.~Rompineve and A.~Westphal, \emph{{Axion Monodromy and the Weak
  Gravity Conjecture}},
  \href{http://dx.doi.org/10.1007/JHEP04(2016)157}{\emph{JHEP} {\bf 04} (2016)
  157}, [\href{http://arxiv.org/abs/1512.03768}{{\tt 1512.03768}}].

\bibitem{Ibanez:2015fcv}
L.~E. Ibanez, M.~Montero, A.~Uranga and I.~Valenzuela, \emph{{Relaxion
  Monodromy and the Weak Gravity Conjecture}},
  \href{http://dx.doi.org/10.1007/JHEP04(2016)020}{\emph{JHEP} {\bf 04} (2016)
  020}, [\href{http://arxiv.org/abs/1512.00025}{{\tt 1512.00025}}].

\bibitem{Resummation}
B.~Heidenreich and J.~Stout, \emph{to appear}.

\bibitem{Erdos09}
L.~Erd\H{o}s, B.~Schlein and H.-T. Yau, \emph{Local semicircle law and complete
  delocalization for wigner random matrices},
  \href{http://dx.doi.org/10.1007/s00220-008-0636-9}{\emph{Communications in
  Mathematical Physics} {\bf 287} (2009) 641--655},
  [\href{http://arxiv.org/abs/0803.0542}{{\tt 0803.0542}}].

\bibitem{TaoVu}
T.~{Tao} and V.~{Vu}, \emph{{Random Covariance Matrices: Universality of Local
  Statistics of Eigenvalues}}, {\emph{Ann. Probab.} {\bf 40} (Dec., 2012)
  1285--1315}, [\href{http://arxiv.org/abs/0912.0966}{{\tt 0912.0966}}].

\bibitem{Bachlechner:2014gfa}
T.~C. Bachlechner, C.~Long and L.~McAllister, \emph{{Planckian Axions in String
  Theory}}, \href{http://dx.doi.org/10.1007/JHEP12(2015)042}{\emph{JHEP} {\bf
  12} (2015) 042}, [\href{http://arxiv.org/abs/1412.1093}{{\tt 1412.1093}}].

\bibitem{Bachlechner:2017hsj}
T.~C. Bachlechner, K.~Eckerle, O.~Janssen and M.~Kleban, \emph{{Systematics of
  Aligned Axions}},
  \href{http://dx.doi.org/10.1007/JHEP11(2017)036}{\emph{JHEP} {\bf 11} (2017)
  036}, [\href{http://arxiv.org/abs/1709.01080}{{\tt 1709.01080}}].

\bibitem{Johnstone:2001wish}
I.~M. Johnstone, \emph{On the distribution of the largest eigenvalue in
  principal components analysis},
  \href{http://dx.doi.org/10.1214/aos/1009210544}{\emph{Ann. Statist.} {\bf 29}
  (04, 2001) 295--327}.

\bibitem{Goodman:1963ddc}
N.~R. Goodman, \emph{The distribution of the determinant of a complex wishart
  distributed matrix},
  \href{http://dx.doi.org/10.1214/aoms/1177704251}{\emph{Ann. Math. Statist.}
  {\bf 34} (03, 1963) 178--180}.

\bibitem{hosotani:1983xw}
Y.~Hosotani, \emph{{Dynamical Mass Generation by Compact Extra Dimensions}},
  \href{http://dx.doi.org/10.1016/0370-2693(83)90170-3}{\emph{Phys.Lett.} {\bf
  B126} (1983) 309}.

\bibitem{bachlechner:2014hsa}
T.~C. Bachlechner, M.~Dias, J.~Frazer and L.~McAllister, \emph{{Chaotic
  inflation with kinetic alignment of axion fields}},
  \href{http://dx.doi.org/10.1103/PhysRevD.91.023520}{\emph{Phys.Rev.} {\bf
  D91} (2015) 023520}, [\href{http://arxiv.org/abs/1404.7496}{{\tt
  1404.7496}}].

\bibitem{Nguyen:2011rml}
H.~H. {Nguyen} and V.~{Vu}, \emph{{Random matrices: Law of the determinant}},
  {\emph{arXiv e-prints} (Dec, 2011) arXiv:1112.0752},
  [\href{http://arxiv.org/abs/1112.0752}{{\tt 1112.0752}}].

\bibitem{Kreuzer:2000xy}
M.~Kreuzer and H.~Skarke, \emph{{Complete Classification of Reflexive Polyhedra
  in Four Dimensions}}, {\emph{Adv.Theor.Math.Phys.} {\bf 4} (2002)
  1209--1230}, [\href{http://arxiv.org/abs/hep-th/0002240}{{\tt
  hep-th/0002240}}].

\bibitem{Witten:1996bn}
E.~Witten, \emph{{Nonperturbative superpotentials in string theory}},
  \href{http://dx.doi.org/10.1016/0550-3213(96)00283-0}{\emph{Nucl.Phys.} {\bf
  B474} (1996) 343--360}, [\href{http://arxiv.org/abs/hep-th/9604030}{{\tt
  hep-th/9604030}}].

\bibitem{Beasley:2005iu}
C.~Beasley and E.~Witten, \emph{{New instanton effects in string theory}},
  \href{http://dx.doi.org/10.1088/1126-6708/2006/02/060}{\emph{JHEP} {\bf 02}
  (2006) 060}, [\href{http://arxiv.org/abs/hep-th/0512039}{{\tt
  hep-th/0512039}}].

\bibitem{Demirtas:2019lfi}
M.~Demirtas, C.~Long, L.~McAllister and M.~Stillman, \emph{{Minimal Surfaces
  and Weak Gravity}},  \href{http://arxiv.org/abs/1906.08262}{{\tt
  1906.08262}}.

\bibitem{Demirtas:2018akl}
M.~Demirtas, C.~Long, L.~McAllister and M.~Stillman, \emph{{The Kreuzer-Skarke
  Axiverse}},  \href{http://arxiv.org/abs/1808.01282}{{\tt 1808.01282}}.

\bibitem{Long:2014fba}
C.~Long, L.~McAllister and P.~McGuirk, \emph{{Heavy Tails in Calabi-Yau Moduli
  Spaces}}, \href{http://dx.doi.org/10.1007/JHEP10(2014)187}{\emph{JHEP} {\bf
  10} (2014) 187}, [\href{http://arxiv.org/abs/1407.0709}{{\tt 1407.0709}}].

\bibitem{Long:2016jvd}
C.~Long, L.~McAllister and J.~Stout, \emph{{Systematics of Axion Inflation in
  Calabi-Yau Hypersurfaces}},
  \href{http://dx.doi.org/10.1007/JHEP02(2017)014}{\emph{JHEP} {\bf 02} (2017)
  014}, [\href{http://arxiv.org/abs/1603.01259}{{\tt 1603.01259}}].

\bibitem{Randall:1998uk}
L.~Randall and R.~Sundrum, \emph{{Out of this world supersymmetry breaking}},
  \href{http://dx.doi.org/10.1016/S0550-3213(99)00359-4}{\emph{Nucl. Phys.}
  {\bf B557} (1999) 79--118}, [\href{http://arxiv.org/abs/hep-th/9810155}{{\tt
  hep-th/9810155}}].

\bibitem{Kachru:2007xp}
S.~Kachru, L.~McAllister and R.~Sundrum, \emph{{Sequestering in String
  Theory}}, \href{http://dx.doi.org/10.1088/1126-6708/2007/10/013}{\emph{JHEP}
  {\bf 10} (2007) 013}, [\href{http://arxiv.org/abs/hep-th/0703105}{{\tt
  hep-th/0703105}}].

\bibitem{Schmaltz:2006qs}
M.~Schmaltz and R.~Sundrum, \emph{{Conformal Sequestering Simplified}},
  \href{http://dx.doi.org/10.1088/1126-6708/2006/11/011}{\emph{JHEP} {\bf 11}
  (2006) 011}, [\href{http://arxiv.org/abs/hep-th/0608051}{{\tt
  hep-th/0608051}}].

\bibitem{Berg:2010ha}
M.~Berg, D.~Marsh, L.~McAllister and E.~Pajer, \emph{{Sequestering in String
  Compactifications}},
  \href{http://dx.doi.org/10.1007/JHEP06(2011)134}{\emph{JHEP} {\bf 06} (2011)
  134}, [\href{http://arxiv.org/abs/1012.1858}{{\tt 1012.1858}}].

\bibitem{Berg:2012aq}
M.~Berg, J.~P. Conlon, D.~Marsh and L.~T. Witkowski, \emph{{Superpotential
  de-sequestering in string models}},
  \href{http://dx.doi.org/10.1007/JHEP02(2013)018}{\emph{JHEP} {\bf 02} (2013)
  018}, [\href{http://arxiv.org/abs/1207.1103}{{\tt 1207.1103}}].

\bibitem{Baumann:2014nda}
D.~Baumann and L.~McAllister, \emph{{Inflation and String Theory}}.
\newblock Cambridge Monographs on Mathematical Physics. Cambridge University
  Press, 2015,
  \href{http://dx.doi.org/10.1017/CBO9781316105733}{10.1017/CBO9781316105733}.

\end{thebibliography}\endgroup
\end{document}